\documentclass[aps,prd,reprint,twocolumn,superscriptaddress,longbibliography,nofootinbib,floatfix,showpacs]{revtex4-1}
\usepackage{epsfig}
\usepackage{amsmath,amssymb,amsfonts}
\usepackage{hyperref}
\usepackage{mathrsfs}
\usepackage{bbm}
\usepackage{slashed}
\usepackage{graphicx}
\usepackage{verbatim}

\usepackage{mathrsfs}

\usepackage[usenames]{color}

\newcommand{\be}{\begin{equation}}
\newcommand{\ee}{\end{equation}}
\newcommand{\bw}{\begin{widetext}}
\newcommand{\ew}{\end{widetext}}
\newcommand{\bi}{\begin{itemize}}
\newcommand{\ei}{\end{itemize}}
\newcommand{\bea}{\begin{eqnarray}}
\newcommand{\eea}{\end{eqnarray}}
\newcommand{\bra}[1]{\langle\,#1\,|}          
\newcommand{\ket}[1]{|\,#1\,\rangle}          
\newcommand{\ud}{\mathrm{d}}

\newcommand{\LCm}{{\scriptscriptstyle -}} 
\newcommand{\LCp}{{\scriptscriptstyle +}}
\newcommand{\LCpm}{{\scriptscriptstyle \pm}}

\newcommand{\LCperp}{{\scriptscriptstyle \perp}}

\newcommand{\vcb}[1]{\mbox{\bf #1}}

\usepackage[T1]{fontenc} \usepackage[latin1]{inputenc}

\DeclareMathAlphabet{\mathpzc}{OT1}{pzc}{m}{it}
\newcommand{\tint}[1]{\int\!\ud{\sf #1}}   
\newcommand{\wavevector}{\kappa}
\newcommand{\fsl}[1]{\slashed{#1}}

\begin{document}

\title{Vacuum refractive indices and helicity flip in strong-field QED}

\author{Victor Dinu}
\email[]{dinu@barutu.fizica.unibuc.ro}
\affiliation{Department of Physics, University of Bucharest, P. O. Box MG-11, M\u agurele 077125, Romania}

\author{Tom Heinzl}
\email[]{theinzl@plymouth.ac.uk}
\affiliation{School of Computing and Mathematics, Plymouth University, Plymouth PL4 8AA, UK}

\author{Anton Ilderton}
\email[]{anton.ilderton@chalmers.se}
\affiliation{Department of Applied Physics, Chalmers University of Technology, SE-41296 Gothenburg, Sweden}

\author{Mattias Marklund}
\email[]{mattias.marklund@chalmers.se}
\affiliation{Department of Applied Physics, Chalmers University of Technology, SE-41296 Gothenburg, Sweden}

\author{Greger Torgrimsson}
\email[]{greger.torgrimsson@chalmers.se}
\affiliation{Department of Applied Physics, Chalmers University of Technology, SE-41296 Gothenburg, Sweden}

\begin{abstract}
Vacuum birefringence is governed by the amplitude for a photon to flip helicity or polarisation state in an external field. Here we calculate the flip and non-flip amplitudes in arbitrary plane wave backgrounds, along with the induced spacetime-dependent refractive indices of the vacuum. We compare the behaviour of the amplitudes in the low energy and high energy regimes, and analyse the impact of pulse shape and energy.  We also provide the first lightfront-QED derivation of the coefficients in the Heisenberg-Euler effective action. 
\end{abstract}

\maketitle
\section{Introduction}

The quantum vacuum, when exposed to intense light, effectively becomes a birefringent medium~\cite{Toll:1952} due to the possibility of light-by-light scattering~\cite{Halpern:1934,Euler:1935zz,Heisenberg:1935qt}. Photonic probes of the vacuum then acquire an ellipticity in their polarisation, in analogy to a probe passing through a birefringent crystal. This `vacuum birefringence' is experimentally accessible due to the advent of high-intensity (optical) and high energy (X-FEL) laser systems. A flagship experiment to measure vacuum birefringence, using combined optical and X-ray lasers, has been proposed by the HIBEF consortium \cite{HIBEF} following~\cite{Heinzl:2006xc}. Continual progress is being made in meeting the experimental challenges~\cite{HP}, in particular in developing the required X-ray polarimetry \cite{Marx:2013}. This paper complements the experimental progress by extending previous theoretical results \cite{Toll:1952,Narozhny:1968,Ritus:1972,Heinzl:2006xc,Shore:2007um} obtained for constant fields to the case of spacetime dependent plane wave fields.

\begin{figure}[t!]
\centering\includegraphics[width=0.7\columnwidth]{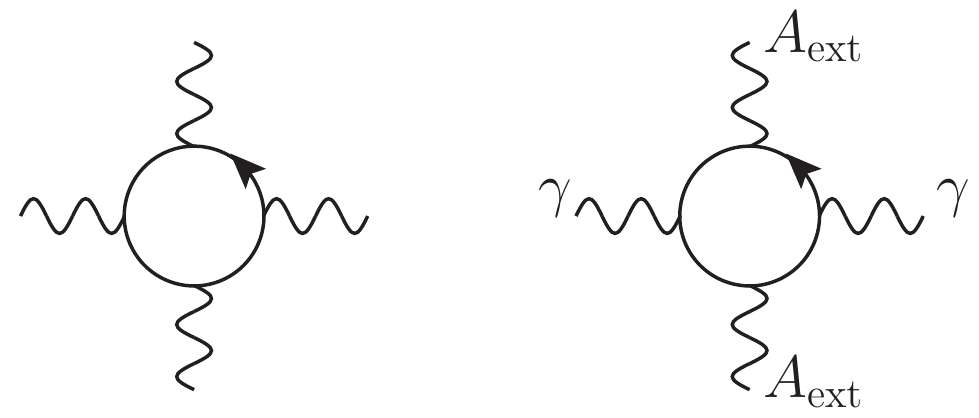}
\caption{\label{FIG:LBL} {\it Left}: the basic photon-photon scattering diagram. {\it Right}: the lowest order contribution to birefringence in a background field. Two of the loop's legs are attached to the background, two to a probe photon.}
\end{figure}

More generally, one can say that it is becoming a feasible agenda to study the vacuum as a nonlinear medium. In this medium, many effects known from classical optics in matter become observable in an entirely photonic setup, i.e.\ the quantum vacuum gives us ``optics without the optics''. Since it is photon scattering which lies behind these effects, they are all captured by the $S$-matrix, see Fig.~\ref{FIG:LBL}~\cite{Euler:1935zz}. For example, vacuum birefringence is part of forward scattering (no momentum change) while backscattering may be viewed as quantum reflection~\cite{Gies:2013yxa}. Off-forward scattering in general corresponds to  diffraction~\cite{Lundstrom:2005za,DiPiazza:2006pr,KING,Davila:2013wba} or deflection \cite{Kim:2011fy}.

Due to laser-laser scattering energies being low compared to the electron rest mass, one normally considers vacuum birefringence from the perspective of the low energy effective Heisenberg-Euler action. From this action one obtains modified Maxwell's equations for the behaviour of a classical electromagnetic field; solving these equations in perturbation theory with a plane wave ansatz one then finds two refractive indices in the vacuum, which lead to e.g.\ a quantum-induced ellipticity in an initially linearly polarised plane wave probe. See~\cite{Shore:2007um,BialynickiBirula:1984tx,Affleck} for lucid treatments along these lines.

The Heisenberg-Euler effective action is the leading term in a derivative expansion, so, strictly, holds only for constant fields. If one is interested in a particular non-constant field geometry, then going beyond leading order typically implies calculating the polarisation tensor $\Pi_{\mu\nu}$ for that field, and from this extracting the refractive indices. See~\cite{Schubert:2000yt,Dittrich:2000zu, Dunne:2004nc} for methods and comprehensive reviews. The polarisation tensor in plane waves was first found in~\cite{Baier:1975ff,Becker-Mitter}. Most recently, a manifestly symmetric representation of this tensor was obtained and the Ward identity verified in~\cite{Meuren:2013oya}. The tensor in homogeneous fields has been reexamined in detail in~\cite{Karbstein:2013ufa}.
 
We will show here that it is not necessary to calculate the whole polarisation tensor in order to obtain the observables relevant for e.g.\ birefringence.  To understand why, recall that a photon of definite momentum and helicity propagates freely in vacuum. The interaction of the photon with other photons, or a background field, leads broadly to three possible effects. First, the photon's momentum is changed. Second, its helicity is flipped. Third, particle number is changed via e.g.\ pair production or photon splitting. For birefringence, one is interested in the case that photons are effectively scattered forward with a change in polarisation, and in the regime where particle production is unlikely (though see~\cite{Uggerhoj}); helicity flip is the physical process underlying the effective birefringence of the vacuum. The relevant observables are built from the helicity-flip and non-flip probabilities which are obtained directly from only two quantum scattering amplitudes. These are the subject of this paper.

Before considering the focussed optical pulses and X-ray probes which will be used in experiments, we want to consider a basic case for which we have full analytic control. We therefore restrict attention here to plane wave backgrounds. (For more complete models of laser pulses see e.g.~\cite{NF,Fedotov}.) This will allow us to explicitly check the applicability of previous constant (crossed) field results; see for example \cite{Toll:1952,Narozhny:1968,Heinzl:2006xc,Shore:2007um,BialynickiBirula:1984tx,Affleck} and references therein.%
  
The paper is organised as follows. We begin below by briefly reviewing the usual approach to vacuum birefringence in strong fields, based on effective actions. In Section~\ref{SECT:NEW} we describe the observables relevant for birefringence and introduce our own approach, based on scattering-amplitudes and expectation values in QED. We show also how to extract the spacetime-dependent vacuum refractive indices for arbitrary plane waves. The probabilities for helicity flip and non-flip are analysed in Sections~\ref{SECT:FLIP} and Section~\ref{SECT:NON} respectively. Since our results are exact we are able to interpolate between the low-energy regime directly relevant to upcoming experiments, and the high-energy regime where the probe energy exceeds the electron mass, $m$. The extension of these results to more realistic, i.e.\ focussed, fields is then discussed in the conclusions.

We use lightfront field theory throughout, as this is the easiest and most natural way to perform QED calculations in plane wave backgrounds~\cite{Neville:1971uc}. The details of this approach are not needed to understand our results (which are equivalent to those obtained via the usual, covariant Feynman diagram approach), and so the explicit calculations are left to the appendices.

\subsection{Preliminaries and Review}\label{SECT:REVIEW}
The standard derivation of vacuum birefringence \cite{Toll:1952,Narozhny:1968,BialynickiBirula:1984tx,Affleck,Heinzl:2006xc,Shore:2007um} starts from the inhomogeneous wave equation (obtained by varying the quantum effective action) in a background field $a$ by linearising in the fluctuating field $b$,
\be  \label{wave-eq}
  \Box b_\mu = j_\mu = \Pi_{\mu\nu}(a) b^\nu \; .
\ee 
Thus, the vacuum current $j_\mu$ is defined in terms of the background dependent polarisation tensor $\Pi_{\mu\nu}(a)$ depicted in Fig.~\ref{FIG:LOOP}. The wave equation (\ref{wave-eq}) is solved by making a geometric optics, or eikonal, ansatz,
\be\label{ansatz}
	b_\mu = \epsilon_\mu \exp \big(-i \Psi\big) \;,
\ee
which implies the algebraic `transport' equation for the polarisation vector $\epsilon$,
\be
  \left\{ \wavevector^2 g_{\mu\nu} + \Pi_{\mu\nu} (\wavevector; a) \right\} \epsilon^\nu = 0 \; ,
\ee
where the probe wavevector $\kappa$ is defined by
\be\label{probe wavevector}
	\wavevector_\mu = \partial_\mu \Psi \;.
\ee
The secular equation then turns into light-cone conditions \cite{Shore:2007um,Gies2} which may be written as
\be \label{LC-cond1}
 	\wavevector^2 + \Pi_\pm = 0 \; ,
\ee
in which the $\Pi_\pm$ are the two nontrivial eigenvalues of the polarisation tensor.

At low photon energies $\ll m$, where $m$ is the electron mass, the appropriate effective Lagrangian is Heisenberg-Euler. To leading order in field strengths this is
\be \label{HE.LO}
  \mathcal{L}_{\mathrm{HE,LO}} = \frac{1}{2} \big( c_- \mathscr{S}^2 + c_+ \mathscr{P}^2 \big) \; .
\ee 
It depends only on the field invariants  
\bea 
  \mathscr{S} &=& -\frac{1}{4} F_{\mu\nu} F^{\mu\nu} = \frac{1}{2} (E^2 - B^2) \label{Sinv} \; , \\
  \mathscr{P} &=& -\frac{1}{4} F_{\mu\nu} \tilde{F}^{\mu\nu} =\vcb{E} \cdot \vcb{B}  \label{Pinv}\; ,
\eea
and the low energy constants are 
\be 
\left\{  \begin{array}{cc}
  	c_+ \\
  	c_-
  \end{array}\right\}
 			=  \frac{\alpha}{45 \pi} \frac{1}{E_S^2}
 	 			\left\{ \begin{array}{cc}
  					7\\
  					4
  				\end{array} \right\}\; .
\ee
In this situation, the eigenvalues of the polarisation tensor may be expressed in terms of the background energy-momentum tensor, $T^{\mu\nu}$,
\be
  \Pi_\pm = c_\pm \wavevector_\mu T^{\mu\nu} \wavevector_\nu \ge 0 \; ,
\ee
which obey a positive energy theorem \cite{Shore:2007um}. The light-cone conditions (\ref{LC-cond1}) then take on the simple form
\be
	\wavevector^2 + \Pi_\pm = (g_{\mu\nu} + c_\pm T_{\mu\nu})\wavevector^\mu \wavevector^\nu = 0 \;,
\ee
with effective metric $g_{\mu\nu} + c_\pm  T_{\mu\nu}$. Introducing an index of refraction
\be\label{n-def}
	\mathrm{n} = \frac{|\underline{\wavevector}|}{\wavevector_0} \;,
\ee
one finds the two solutions
\be \label{N.PM1}
  \mathrm{n}_\pm = 1 + c_\pm \, \frac{\wavevector_\mu T^{\mu\nu} \wavevector_\nu}{2\omega^2}\ \to 1 +  \frac{\alpha}{45 \pi}
  (11 \pm 3) \frac{E^2}{E_S^2} \; ,
\ee
where the final expression is obtained for head-on collisions between the probe and background. One can obtain (\ref{HE.LO}) and other effective Lagrangians by simply writing down all possible invariant combinations of the field strength, with unknown coefficients. The `matching problem' is then to calculate these coefficients in the underlying theory. QED, of course, implies the Heisenberg-Euler Lagrangian (\ref{HE.LO}), while the low-energy limit of the U(1) sector in string theories tends to be Born-Infeld electrodynamics~\cite{BI,BialynickiBirula:1984tx,ZZ}. In this paper we will give, to the best of our knowledge, the first calculation of the matching coefficients $c_\pm$ in lightfront QED.
\section{Observables for birefringence} \label{SECT:NEW}
The proposed setup for measuring birefringence uses an intense optical laser probed by an X-FEL beam. Both are linearly polarised. As the probe, with polarisation $\epsilon$, passes through the optical laser, it acquires an ellipticity due to light-by-light scattering, and it is this ellipticity which is to be measured.

The observables of interest are then the polarisation of the probe field, or the number of probe photons detected with a nonzero $\epsilon'$--polarisation component, orthogonal\footnote{This does not mean the beam's polarisation is rotated by $90^\circ$. Rather, it is the presence of a small number of orthogonally-polarised photons which gives the beam a small ellipticity.} to that of the initial photons, $\epsilon'\epsilon=0$. Such observables can be obtained in quantum field theory as the expectation values of (i) the $\epsilon'$--projected photon number operator, $N_{\epsilon'}$, and (ii) the $\epsilon'E$ component of the laser field, both calculated in the time-evolved state.

\begin{figure}[t!!]
\centering\includegraphics[width=0.7\columnwidth]{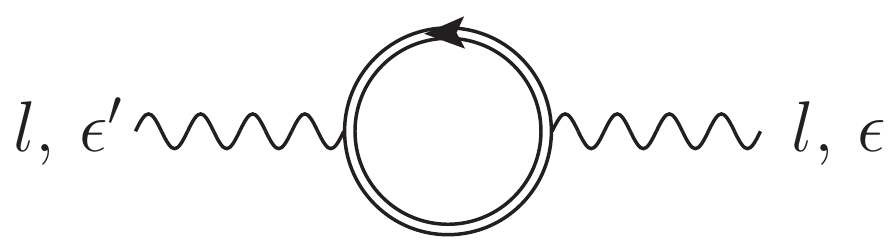}
\caption{\label{FIG:LOOP} The one-loop contribution to helicity or polarisation flip in a background field. The tree-level contribution is identically zero.}
\end{figure}

We describe the background field as a plane wave which depends on $x^\LCp := t + z$, which is `lightfront time'. As such, our system has an explicit dependence on $x^\LCp$, which is the natural parameter to use in order to describe time evolution~\cite{Neville:1971uc}. A laser probe is best described by an asymptotic coherent state. Such `closest to classical' states give, for example, nonzero expectation values for electromagnetic fields, unlike pure photon--number states. The expectation value for an observable $O$ in the time evolved state is
\be\label{snitt}
	\langle O \rangle_\text{out} = \bra{\text{in}}U^\dagger(x^\LCp) O U(x^\LCp) \ket{\text{in}} \;,
\ee
where $U(x^\LCp)$ is the time-evolution operator (so that $S:=U(\infty)$ is the S-matrix operator) and $\ket{\text{in}}$ is a suitable coherent state. Working to one loop, we find that the observables introduced above can all be expressed in terms of a single lightfront time-ordered transition amplitude $T$, defined by sandwiching the time evolution operator between single photon states $\ket{l,\epsilon}$ (for momentum $l_\mu$ and polarisation $\epsilon$):
\be\label{TDEF}
	\bra{l',\epsilon'}U(x^\LCp)\ket{l,\epsilon} =: \delta({\sf l}',{\sf l})(-\epsilon'\epsilon+iT_{\epsilon'\epsilon}(x^\LCp)) \;.
\ee
The relativistically invariant three--dimensional delta function $\delta$ appears because of the conservation properties in a plane wave background. The factor of $i$ is for convenience, and it is understood throughout that $T$ also depends on the forward--scattered probe momentum $l_\mu$. The relevant one-loop Feynman diagram is given in Fig.~\ref{FIG:LOOP}, and the double lines are dressed fermions (Volkov~\cite{Volkov}) propagators. While the method of calculation, starting from this diagram, is by now standard, it is more efficient and elegant to calculate in lightfront field theory. The derivation and renormalisation of $T$ is given in Appendix~\ref{APP-A}. Here we describe how $T$ enters the observables of interest.

We take the incoming probe field to be monochromatic, to illustrate. The coherent state for such a probe is defined by the action of the photon annihilation operator,
\be
	a_\mu(l')\ket{\text{in}}=-i\frac{E_0}{2l_0}\epsilon_\mu {\delta}({\sf l'},{\sf l})\ket{\text{in}} \;.
\ee
Before interaction with the background field, the fields of this state are $\epsilon$-polarised, so that
\be
	\epsilon \langle E\rangle_\text{in} = E_0 \cos lx \;, \quad \epsilon'\langle E\rangle_\text{in} = 0\;,
\ee
where $\epsilon'$ is the second possible polarisation of the wave, $\epsilon'\epsilon=0$. Interaction with the background changes the structure of the fields. First, the fields acquire an ellipticity, because the interaction generates a nonzero $\epsilon'$ field component
\be\label{ell1}
	\epsilon'\langle E\rangle_\text{out} = \text{Re } E_0 e^{-ilx} iT_{\epsilon'\epsilon}(x^\LCp) \;.
\ee
The original field component is also affected, becoming\footnote{Here we have neglected terms of higher order in the probe field strength $E_0$. These terms vanish for e.g.\ (\ref{ell1}) when $\epsilon'\epsilon=0$.}
\be\label{evolve}
\begin{split}
\epsilon\langle E\rangle_\text{out} &= \text{Re } E_0 e^{-ilx}\big(1+ iT_{\epsilon\epsilon}(x^\LCp)\big) \\
	&\simeq \text{Re}\, E_0 e^{-ilx + iT_{\epsilon\epsilon}(x^\LCp)} \;.
\end{split}
\ee
As $T$ is complex, we will have both dispersive and absorptive contributions. As will shortly be shown, though, $T$ is real in the low-energy limit, in which case the induced field component (\ref{ell1}) becomes $\langle\epsilon'E\rangle = E_0\sin(lx)T_{\epsilon'\epsilon}(x^\LCp)$, out of phase with the original field component. The induced ellipticity, $\psi$, of the field is the ratio of the field's minor and major axes, i.e.\ what remains after dropping the trig terms,
\be\label{psi}
	\psi = T_{\epsilon'\epsilon} \;.
\ee
From (\ref{evolve}) we can use (\ref{ansatz}) to read off a (complex) phase shift $\Psi$,
\be\label{evolve2}
	\Psi =lx -T_{\epsilon\epsilon}(x^\LCp) \;,
\ee
and from (\ref{probe wavevector}) we identify the probe wavevector $\kappa$,
\be
	\wavevector_\mu = \partial_\mu \Psi =  l_\mu -n_\mu \frac{\partial T_{\epsilon\epsilon}(x^\LCp)}{\partial x^\LCp} \;.
\ee
From this the refractive index is extracted as in (\ref{n-def}),
\be\label{F1-n}
	\mathrm{n}(x^\LCp) = \frac{|{\underline\wavevector}|}{\wavevector_0}  = 1 +\frac{nl}{\omega_l^2}\frac{\partial T_{\epsilon\epsilon}(x^\LCp)}{\partial x^\LCp} \;.
\ee
Note that the indices are not constant, but depend on $x^\LCp$. Measurements in a real experiment will of course be made far from interaction volumes, and so the asymptotic limit, $x^\LCp\to\infty$, is relevant. In this limit, it is clear from the definition (\ref{TDEF}) that $iT$ is just the one-loop $S$-matrix element for forward scattering between different possible polarisation states. Now, a photon has only two independent polarisation states, so we can take $\{\epsilon', \epsilon\}$ as a basis. (One such choice, see below, is a helicity basis.) The probability that the photon `flips' polarisation between these states during its interaction with the background is just the large time limit of $T_{\epsilon'\epsilon}$ mod-squared, which governs the induced ellipticity in the beam,
\be
	\mathbb{P}_l(\text{flip}) = |	T_{\epsilon'\epsilon}(\infty)|^2 \;.
\ee
Further, $iT_{\epsilon\epsilon}$ is just the non-flip scattering amplitude, and if the number density of $\epsilon$--polarised photons in the in-state is $\rho_\epsilon(l)$ (going beyond plane-wave probes) it follows that the number of orthogonally polarised photons which can be detected in the out-state is
\be\label{N}
	\langle N_{\epsilon'} \rangle_\text{out} = \int\!\ud{\sf l}\ \rho_\epsilon(l) \, \mathbb{P}_l(\text{flip}) = 	\int\!\ud{\sf l}\ \rho_\epsilon(l) \, |T_{\epsilon'\epsilon}(\infty)|^2 \;,
\ee
where $\ud{\sf l}$ is the invariant on-shell momentum measure.

\subsection{Scattering and the polarisation tensor}
Again from (\ref{TDEF}), the asymptotic limit $T(\infty)$ can be written as the polarisation tensor sandwiched between different polarisation states,
\be
	T_{\epsilon_2\epsilon_1}(\infty) = \epsilon_2\Pi(l,l)\epsilon_1 \;,
\ee
for $l_\mu$ the forward-scattered momentum. Given any (two--component) basis $\epsilon^\mu_\lambda$ of the photon's physical polarisations, and writing $T_{\lambda\lambda'}=T_{\epsilon_\lambda \epsilon_{\lambda'}}$ for compactness, the plane-wave polarisation tensor for on-shell photons can be reconstructed as
\be\label{TENSOR}
	\Pi^{\mu\nu} = \sum\limits_{\lambda\lambda'} \epsilon_\lambda^\mu\,  T_{\lambda\lambda'} \, \bar{\epsilon}^\nu_{\lambda'} \;.
\ee
Note, though, that to calculate the observables (\ref{ell1}), (\ref{psi}) and (\ref{N}), one needs only the flip amplitude. This single amplitude corresponds to a particular component of the polarisation tensor, but the tensor as a whole is not needed.

The expressions we have derived for $T$ are simple and easily evaluated in comparison to the majority in the literature. To give these expressions, we begin by introducing our conventions and notation in more detail. Consider a photon probing a background plane wave. The plane wave is a transverse function of $nx$ where $n_\mu$ is a lightlike vector, $n^2=0$. We can choose $n_\mu$ such that $nx = x^0+x^3 \equiv x^\LCp$, lightfront time. (The remaining coordinates are $x^{\LCm}=x^0-x^3$, $x^\LCperp=\{x^1,x^2\}$, $p_\LCpm=(p_0\pm p_3)/2$, $p_\LCperp=\{p_1,p_2\}$.) In order to make approximations and estimates, we introduce $\omega$, a typical frequency scale associated with the background. We will often use $\phi:= kx$, for $k=\omega n$, as a dimensionless lightfront time variable.  The background plane wave is
\be\label{F}
	eF^\text{ext}_{\mu\nu}(\phi) = k_\mu a'_\nu(\phi)-a'_\mu(\phi)k_\nu\;,  \quad a'_\LCperp(\phi) := -e E_\LCperp(\phi)/\omega \;.
\ee
A photon with momentum $l_\mu$ has two independent physical polarisation states. These are described by a two-component basis of polarisation vectors $\epsilon_\lambda^\mu (l)$ which, in the lightfront form, can be chosen to be orthogonal to {\it both} the photon's momentum and to the lightfront/laser direction $n_\mu$, so $l\epsilon_\lambda(l)= n\epsilon_\lambda(l) = 0 \;.$ With vector components in order $+,-,\perp$ we have
\be\label{vector-form}
\epsilon_\lambda^\mu(l)= (0, -\frac{l_\LCperp\varepsilon_\lambda^\LCperp}{l_\LCm},\varepsilon^\LCperp_\lambda)
\ee
The polarisation vectors (which may now be complex) obey the completeness relation
\be
	\sum\limits_{\lambda} {\bar\epsilon}^\mu_\lambda(l) \epsilon^\nu_\lambda(l) = -g^{\mu\nu} + \frac{n^\mu l^\nu + l^\mu n^\nu}{nl} \;.
\ee
Possible choices of bases are the helicity basis
\be\label{circular polarisation vectors}
		\varepsilon^\LCperp_\lambda= \frac{1}{\sqrt{2}}(1,i\lambda) \;,  \qquad \lambda = \pm 1 \;,
\ee
or a basis of linear polarisations,
\be\label{linear polarisation vectors}
		\varepsilon^\LCperp_1=(1,0) \qquad \varepsilon^\LCperp_2=(0,1) \;.
\ee
We will consider both choices below. For the presentation of our general result we let $\epsilon$, $\epsilon'$ be any two polarisation vectors constructed from such a basis. The transition amplitude $T$ depends on the background field through the moving average
\be\label{average}
	\langle a_\LCperp\rangle := \frac{1}{\theta}\int\limits^{\phi+\tfrac{1}{2}\theta}_{\phi-\tfrac{1}{2}\theta}\!\ud x\ a_\LCperp(x) \;,
\ee
where $\phi$ and $\theta$ are lightfront times which arise in the $S$-matrix calculation as centre-of-mass and relative co-ordinates, respectively. The amplitude also depends on the projection of the average (\ref{average}) onto the polarisation vectors, $A:= \epsilon \langle a \rangle$ and $\bar{A} := {\bar\epsilon}' \langle a \rangle$. We write a subscript $\theta$ or $\phi$ for a derivative, so $A_\theta \equiv \epsilon^\mu \partial_\theta \langle a_\mu \rangle$ and so on. The asymptotic amplitude $T(\infty)$ is
\bw
\be\label{T-resultat}
\begin{split}
T_{\epsilon'\epsilon}(\infty)=-\frac{\alpha}{\pi}\frac{1}{kl}\int\limits_{-\infty}^\infty\!\ud \phi \!\int\limits_0^\infty\!\ud\theta\theta\  \bigg( \mathcal{I}_1\big(\tfrac{\theta M^2}{kl}\big)&\bigg[\frac{\bar{\epsilon}'\epsilon}{2} \frac{M^2-m^2}{\theta^2M^2} \frac{\partial (\theta M^2)}{\partial\theta} +(\bar{A}_\theta+\tfrac{1}{2}\bar{A}_\phi)(A_\theta-\tfrac{1}{2}A_\phi) \bigg] \\
 + &\frac{1}{2}\mathcal{I}_2\big(\tfrac{\theta M^2}{kl}\big)\bigg[\frac{\bar{\epsilon}'\epsilon }{2}\langle a'\rangle^2 -(\bar{A}_\phi A_\theta-\bar{A}_\theta A_\phi)\bigg] \bigg)
\end{split}
\ee
\ew
in which 
\be
	M^2(\phi,\theta) = m^2+\langle a \rangle^2-\langle a^2\rangle \;
\ee
is Kibble's effective mass~\cite{Kibble:1975vz}, and the two $\mathcal{I}$-functions are simple combinations of modified Bessel functions,
\be\begin{split}
	\mathcal{I}_1(x) &= ix e^{-ix} \big(K_1(ix)- K_0(ix)\big) \;, \\
	\mathcal{I}_2(x) &= e^{-ix} K_0(ix) \;.
\end{split}
\ee
The loop momentum implicit in Fig.~\ref{FIG:LOOP} has been integrated out exactly, which is one reason why this amplitude has a much simpler representation than is typically obtained for the polarisation tensor. The two remaining integrals over the lightfront times $\{\phi,\theta\}$, which come from the two vertices in Fig.~\ref{FIG:LOOP}, can be performed numerically. This is discussed in Appendix~\ref{APP-C}. Despite appearances, the integrals are well behaved at $\theta=0$; a potential contact term~\cite{Dittrich:2000zu} (a UV divergence) is removed by renormalisation, see Appendix~\ref{APP-A}.

The remainder of the paper is given over to a detailed investigation of the cases of interest, namely the flip and non-flip amplitudes. In order to cement the fact that $T$ is the relevant object, we will `preview' the discussion with $T$'s low energy expansion, which is relevant for upcoming birefringence experiments. Let $l_\mu$ be the probe momentum and $k_\mu$ the typical momentum associated with the background, as above. A low energy limit can then be defined by expanding in the invariant $kl/m^2$. Although we have presented $T(\infty)$ above, the low energy part of $T(x^\LCp)$ has the same form, except for the integration limits, and is equal to
\be
\label{Low energy R}
	T_{\epsilon'\epsilon}(x^\LCp)=\frac{\alpha}{4\pi}\frac{kl}{m^4}\frac{1}{45}\int\limits^{kx}_{-\infty}\!\ud\varphi\ (14\bar{\epsilon}'\epsilon a'^2-6\bar{\epsilon}'a'\epsilon a' ) \;.
\ee
We immediately recognise the coefficients for the refractive indices and ellipticity. For non-flip, $\epsilon'=\epsilon$, there are two independent cases to consider; if $\epsilon \mathbf{E}=0$ we pick up the first term in (\ref{Low energy R}) with coefficient $14$, and if $\epsilon \parallel \mathbf{E}$ we pick up both terms, $14-6=8$. Using (\ref{F1-n}), these give the coefficients of the two refractive indices. For the induced ellipticity (\ref{psi}) and its asymptotic limit  $T_{\epsilon'\epsilon}(\infty)$ with $\bar{\epsilon}'\epsilon=0$, i.e.\ the flip probability, we pick up only the last term in (\ref{Low energy R}), with coefficient $6$. This is the difference of the two refractive indices, and it is this which is measured by a birefringence experiment.

\section{Helicity and polarisation flip}\label{SECT:FLIP}
In this section we analyse the flip probability. We will frame the discussion in terms of helicity, which is most natural from a high-energy physics perspective since a photon's helicity is relativistically invariant~\cite{Weinberg:1995mt}, and preserved in the absence of interactions. Our calculations hold, though, in arbitrary bases.

The normalised photon state of definite helicity $\lambda$ and momentum wavepacket $f$ is
\be\label{foton-in}
	\ket{f,\lambda}= \int\!\ud{\sf l}' f(l')\ \epsilon_\lambda(l') {\hat a}^\dagger(l')\ket{0} \;, \quad \int\!\ud{\sf l}' |f(l')|^2=1 \;.
\ee
In the presence of a background, there is a nonzero probability that the photon's helicity will flip, $\lambda\to -\lambda$, as it propagates. In general backgrounds this can be accompanied by scattering, but when the background is a plane wave, conservation of three lightfront momentum components means that asymptotic scattering is automatically forward. (See \cite{Affleck} for non-forward, non-asymptotic contributions using the effective action approach.) Hence the  probability for helicity flip $\lambda\to-\lambda$ and forward scattering is here given by the total probability of helicity flip,
\be
	\mathbb{P}(\text{flip}) =  \int\!\ud{\sf l}' \big|\bra{l',-\lambda} S \ket{f,\lambda} \big|^2 \;.
\ee
Taking the wavepacket to be peaked around momentum $l_\mu$, the flip probability becomes
\be\label{PFLIP}
	\mathbb{P}_l(\text{flip}) = |T_{\epsilon'\epsilon}(\infty)|^2 \;,\quad \bar{\epsilon}'\epsilon=0 \;.
\ee
In this case only the second terms in each square bracket of (\ref{T-resultat}) contribute.  The probability vanishes when the field vanishes, as it should.
\subsection{Helicity flip: behaviour as a function of energy}
\begin{figure}[t!]
	\includegraphics[width=\columnwidth]{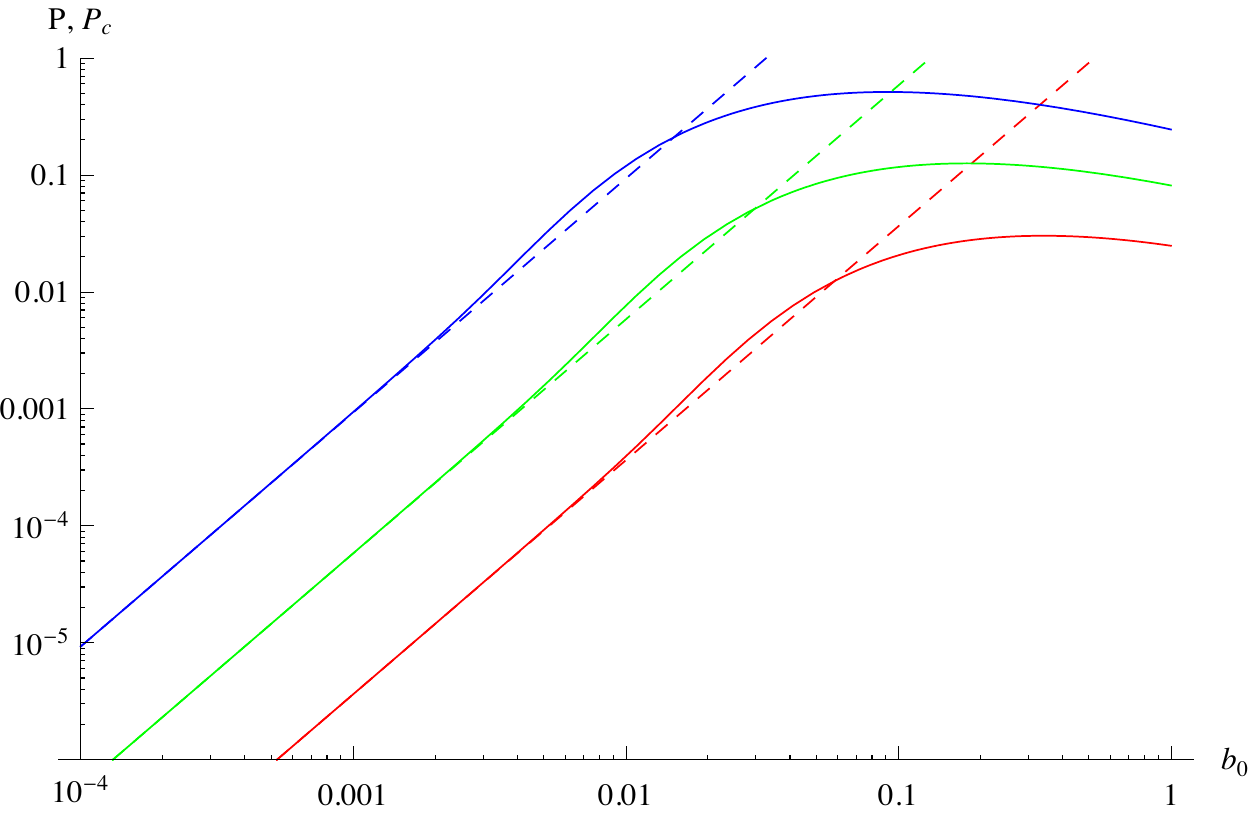}
	\includegraphics[width=\columnwidth]{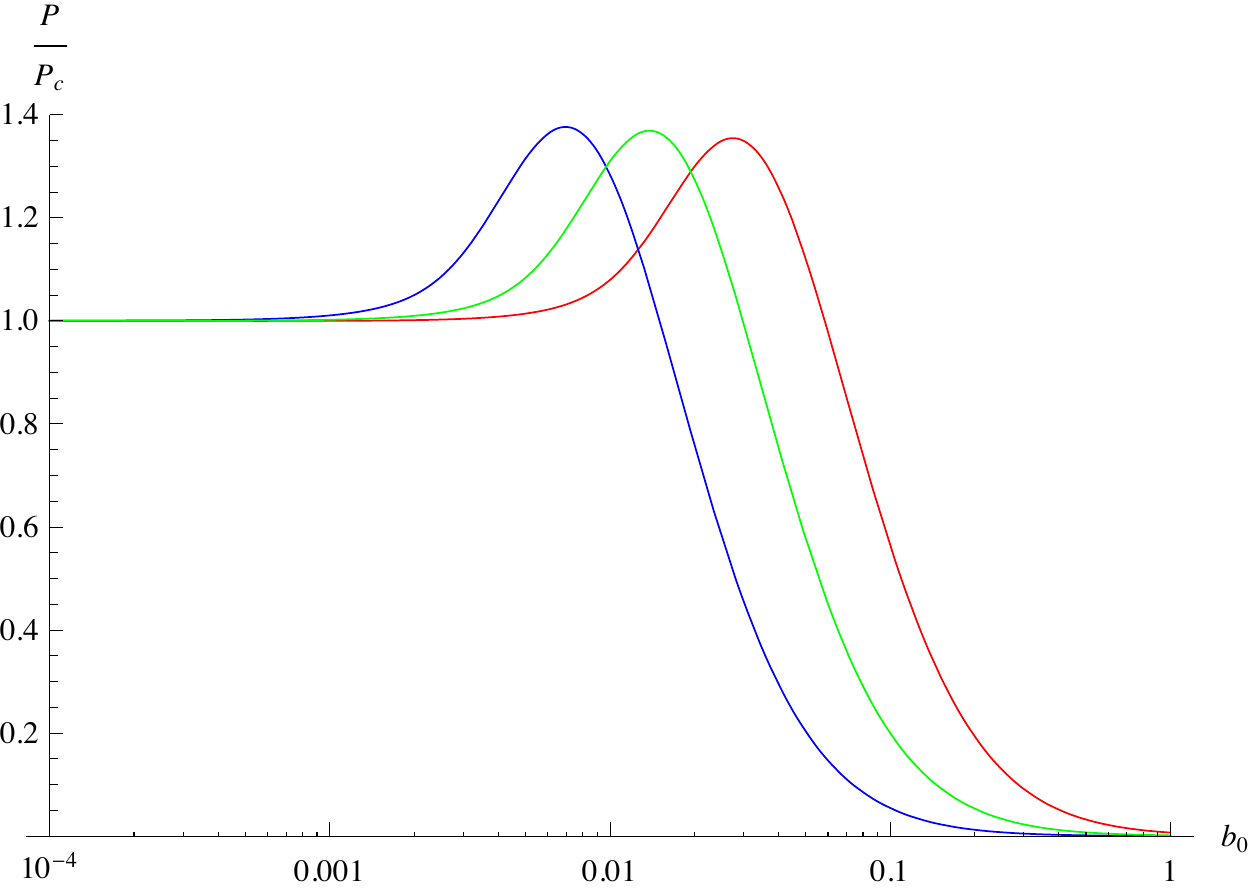}
	\caption{\label{FIG:FLIP} Upper panel: The exact probability $P$ for helicity flip $\lambda\to-\lambda$ (full lines), from (\ref{PFLIP}), and the approximation $P_c$ from (\ref{momentum expansion}) (dashed lines), plotted as functions of invariant $b_0=kl/m^2$ (log-log scale). Lower panel: Ratio $P/P_c$ as a function of $b_0$. The background is circularly polarised with a Gaussian profile, see (\ref{profile}), width $\Lambda=10\pi$ and $a_0=200,~100,~50$ (left to right). In the low energy regime (\ref{momentum expansion}) provides an  excellent approximation.}
\end{figure}

For laser-based investigations of birefringence, we are mainly interested in low-energy (compared to the electron rest mass) probes in low-energy backgrounds, which means that the invariant $b_0:= kl/m^2\ll 1$ and we can simplify the probability. The most naive way of doing so is to change variables $\theta\to b_0\theta$ and then expand in powers of $b_0$. The $\theta$ integral then becomes simple, and we are left with
\be\label{momentum expansion}
	\mathbb{P}_l(\text{flip}) \overset{b_0\ll1}{=}\left| \frac{\alpha}{30\pi}\frac{kl}{m^4} \int\!\ud\phi\ \bar{\epsilon}a'(\phi)\epsilon a'(\phi)\right|^2\;.
\ee
In Fig.~\ref{FIG:FLIP} we compare (\ref{momentum expansion}) with the exact polarisation flip probability (\ref{PFLIP}). For our plots, here and in subsequent examples, we use the following background field profile:
\be\label{profile}
	\bigg(\begin{array}{c}
	a_1 \\
	a_2 
	\end{array}\bigg) 
	= a_0 m e^{-\phi^2/\Lambda^2} \bigg(\begin{array}{c}
 				\cos\varphi \sin(\phi-\phi_{01}) \\
				\sin\varphi \sin(\phi-\phi_{02})
				\end{array}
		\bigg) \;,
\ee
which has an amplitude $\sim a_0$, a Gaussian envelope, and a polarisation determined by $\varphi$ and the carrier phases $\phi_{0i}$. For Fig.~\ref{FIG:FLIP} we have chosen a circularly polarised background, $\varphi=\pi/4$, $\phi_{01}=0$, $\phi_{02}=\pi/2$, and $\Lambda=10\pi$ which corresponds to a FWHM duration of around $30$~fs at optical frequency $\omega=1.24$~eV. We see first that the approximation (\ref{momentum expansion}) is excellent in the low energy regime of interest. As the probe energy rises, (\ref{momentum expansion}) underestimates the full result, over a range of around one order of magnitude in $b_0$. Thus increasing the energy boosts the flip probability. For even higher energies, (\ref{momentum expansion}) greatly overestimates the true probability, which begins to drop. The reason is that at high energies, additional quantum effects appear. There is for example a greater probability that the photon will `decay' via e.g.\ stimulated pair production~\cite{Nikishov:1963, Nikishov:1964a, Schutzhold:2008pz, Heinzl:2010vg,Nousch:2012xe,Titov:2012rd}, and it follows from unitarity that `photon persistence' probabilities must drop. As we will see below, this is associated to the appearance, at high-energy, of effects such as anomalous dispersion, which are not captured by the low-energy approximation.

From the figures, we can read off when the approximation (\ref{momentum expansion}) holds. We find that this can be phrased in terms of a single parameter, namely the photon's ``quantum efficiency parameter'' $\chi$~\cite{Nikishov:1963, Nikishov:1964a, DiPiazza:2011tq}, which is the contraction of the probe momentum with the energy momentum tensor in a plane wave~\cite{Shore:2007um, Heinzl:2006xc},
\be
	\chi = \sqrt{\frac{e^2}{m^6} l^\mu T_{\mu\nu} l^\nu}  \sim \frac{E}{E_S} \frac{\omega_l}{m}  \;.
\ee
Here and below, $\omega_l = l_0$ is the probe energy. $\chi$ is spacetime dependent. As we will see below, low-energy processes are typically sensitive to the total, integrated, $\chi$, but since we have fixed pulse shape and length, peak and total $\chi$ are proportional, and it is simpler to use peak $\chi$, which for the circular polarised wave is
\be
	\chi \stackrel{\text{peak}}{=} \frac{1}{\sqrt{2}}a_0b_0 \;,
\ee
Values of $\chi$ cannot be read off directly from the plots, but are easily  extracted from the numerics. We find that the difference between (\ref{PFLIP}) and (\ref{momentum expansion}) is within 5\% for $\chi< 0.3$, and within 10\% for
\be
	\chi <0.4 \;.
\ee
Even for an optical fieldstrength of $E=10^{-4}E_S$, corresponding to an intensity of around $10^{23}$ W/cm$^2$, the approximation (\ref{momentum expansion}) remains valid for photon energies reaching $\omega_l \sim 1$~GeV. The worst {\it underestimate}, (\ref{momentum expansion})/(\ref{PFLIP}) $\sim 0.72$,  occurs at $\chi\simeq 1$. The probability for helicity flip is maximal for $\chi\sim 12$. This is all consistent with expectations, which are that quantum effects become more important as $\chi$ increases toward unity, although our results allow us to be a little more specific. Interestingly, Fig.~\ref{FIG:FLIP} shows that (\ref{momentum expansion}) again becomes a good approximation even at some higher energies for which $\chi>1$. This is due to the behaviour of the probability, which turns and begins to drop, and so the low-energy expansion changes from an under- to an overestimate. There is then a small kinematic regime around $\chi\simeq 2$ in which the approximation becomes close to the true probability again. It may be for such high probe energies, higher loops are needed in~(\ref{PFLIP}).

\subsection{Locally constant approximation}
More about the low-energy approximation, and its shortcomings, is revealed  by considering the locally constant approximation to (\ref{PFLIP}). This can be obtained either by expanding (\ref{PFLIP}) in the relative coordinate $\theta$, or by writing down (\ref{PFLIP}) for a constant crossed field and then replacing ${\bf E}\to \bf{E}(\phi)$ everywhere. For the latter method, take $\omega=m$, and define $\mathcal{E}^j= {\bf E}^j/E_S$  ($\mathcal{E}^0=0$), the ratio of the electric field to the Schwinger field. A vector potential for the crossed field is $a^\LCperp(\phi) = -m \mathcal{E}^\LCperp \phi$. Representing the Bessel functions as an integral over dimensionless $s$ (which can, see Appendix~\ref{APP-A}, be interpreted as a lightcone momentum fraction) the constant field probability is
\be
	\mathbb{P}_\text{CF}=\bigg| \frac{\alpha m^2}{\pi kl} \!\! \int\limits_{-\infty}^\infty\!\ud \phi\! \int\limits_0^\infty\!\ud\theta \theta \!\int\limits_0^1\! \ud s \ e^{-i\frac{m^2}{kl}\frac{\theta - \mathcal{E}^2\theta^3/12}{2s(1-s)}} 
	\tfrac{1}{4}(\epsilon\mathcal{E})(\bar{\epsilon}\mathcal{E}) \bigg|^2.
\ee
This diverges quadratically with the volume of lightfront time $\phi$. However if we replace $\mathcal{E}\to \mathcal{E}(\phi)$ then we obtain the locally constant approximation, $\mathbb{P}_\text{LCFA}$. In this case we can perform the $\theta$-integral exactly, obtaining Airy ($\mathrm{Ai}$) and Scorer ($\mathrm{Gi}$) functions, both of which are typical of one-loop results in constant fields, see \cite{Heinzl:2006pn,Shore:2007um},
\be\begin{split}
	\mathbb{P}_\text{LCFA}=\bigg|&\frac{\alpha kl}{m^2} \int\!\ud\phi\ \bar\epsilon\mathcal{E}(\phi){\epsilon}\mathcal{E}(\phi) \times \\
	&\int\!\ud s[s(1-s)]^2 \mu^2 \frac{\ud}{\ud\mu}\big[ \mathrm{Ai}(\mu)-i\mathrm{Gi}(\mu)\big]\bigg|^2 \;,
\end{split}
\ee
where
\be
	\mu=\left(\frac{m^2}{kl}\frac{1}{s(1-s)}\right)^{2/3}(-2\mathcal{E}^2)^{-1/3} \;.
\ee
The kinematic regime of experimental interest is $kl/m^2 \ll 1 \iff \mu \gg1$; we therefore turn to the asymptotic expansions of the Airy and Scorer functions, which for $\mu\gg 1$ are
\be
	\mathrm{Ai}'(\mu)=-\frac{\mu^{1/4}}{2\sqrt{\pi}}\exp-\frac{2}{3}\mu^{3/2} \;, \quad \mathrm{Gi}'(\mu)=-\frac{1}{\pi\mu^2} \;.
\ee
The Airy function term is {\it nonperturbatively small} in the invariant $b_0=kl/m^2$, 
relative to the Scorer function. The locally-constant approximation reveals that the helicity flip probability contains a part which is nonperturbative in $b_0$, and which is, naturally, missed when one simply expands in powers of $b_0$. Nevertheless, since this nonperturbative part is exponentially small in the regime of interest, it is safe to neglect it; doing so, the remaining $s$-integral is trivial and we recover precisely (\ref{momentum expansion}), so $\mathbb{P}_\text{LCFA} = \mathbb{P}(\text{flip})$ for $b_0\ll1$. The validity of this approximation in the regime of interest is confirmed by the excellent agreement between (\ref{momentum expansion}) and the exact result (\ref{PFLIP}), shown in Fig.~\ref{FIG:FLIP} and Fig.~\ref{FIG:P-vs-P}.

We have therefore seen, either by direct low-energy expansion or by going via the locally-constant approximation, that the flip probability collapses to a simple expression which is quadratic in the background, i.e.\ is the same as would be obtained to lowest order by treating the background perturbatively. This is as expected from the low-energy expansion of the Heisenberg-Euler action, and is an explicit example of the general statement in~\cite{DiPiazza:2013iwa} that higher-order terms (in the background) are hard to observe in loop processes when probes have low-energy. 

\subsection{Polarisation flip: dependence on geometry}
Since (\ref{PFLIP}) is not restricted to just the helicity states, we consider now the `polarisation flip' probability between elements of the linear basis (\ref{linear polarisation vectors}). This is relevant because the experimental case of interest, birefringence, takes both the background and probe to be linearly polarised. Let us then make the connection with literature results on birefringence.

\begin{figure}[t!]
	\includegraphics[width=\columnwidth]{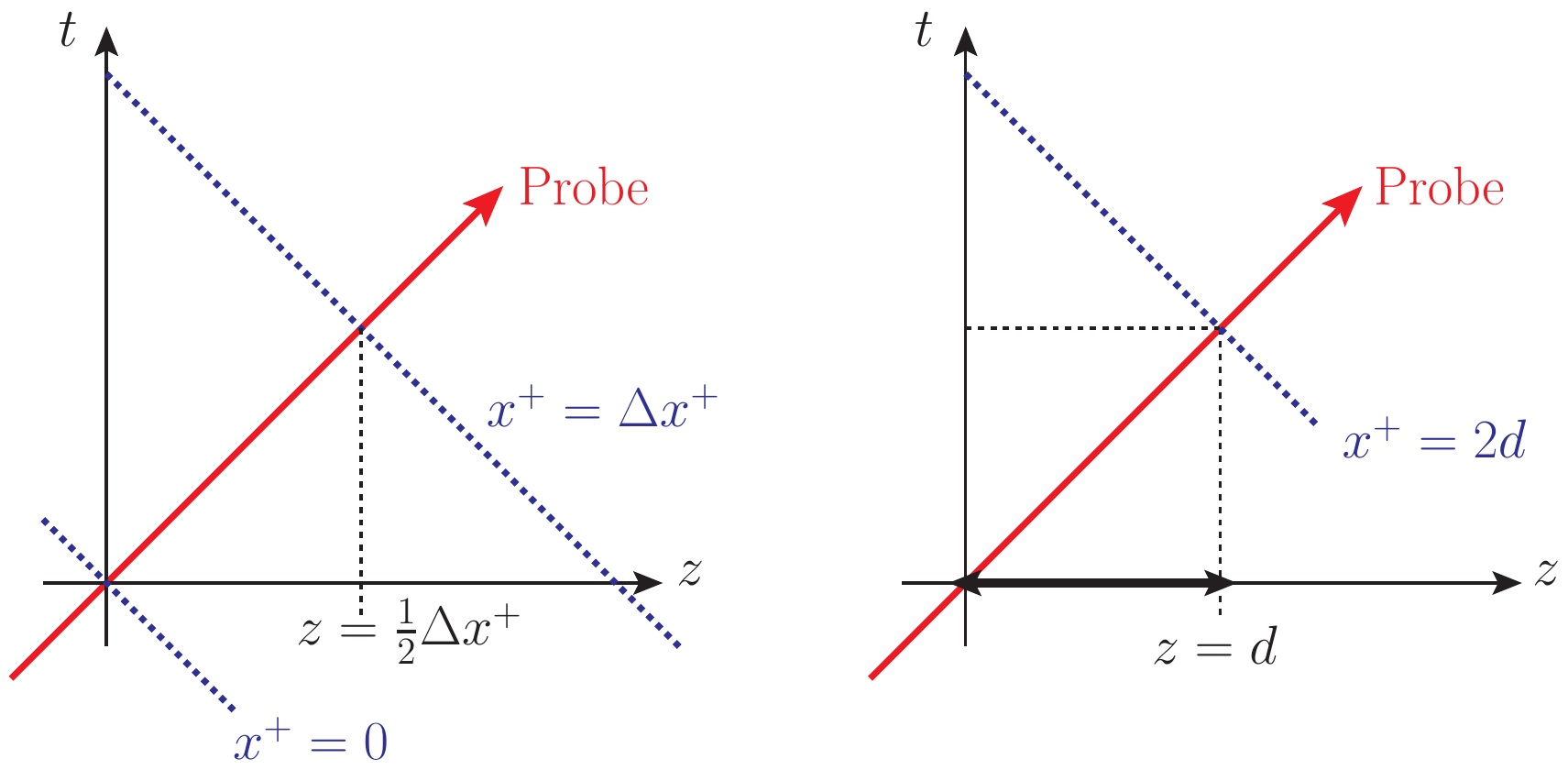}
	\caption{\label{rumtid} Head-on collision between the probe and the background field. The background field has lightfront duration $\Delta x^\LCp$, and the probe emerges after a distance $d=\Delta x^\LCp$/2.}
\end{figure}

We take the incoming photon to have polarisation $\epsilon=\epsilon_1$ from (\ref{vector-form}) and (\ref{linear polarisation vectors}), and we look at the probability that this photon will have flipped to the orthogonal linear polarisation $\epsilon' = \epsilon_2$ after interaction with the background. We begin in the low-energy regime. From (\ref{momentum expansion}) we can then identify three parameters on which the probability depends; the collision angle, the angle between the polarisations of the background and probe, and (somewhat broadly) the background field profile. Since the polarisation vectors are real in this linear setup, we can look directly at the flip amplitude $T$ which we know from (\ref{psi}) is the induced ellipticity of the probe beam. Let $\vartheta$ be the angle between the beam directions $\bf{k}$ and $\bf{l}$, and let $\sigma$ be the angle between $\epsilon_\LCperp$ and $a_\LCperp$. Then,
\be\label{S}
	T_{\epsilon'\epsilon} = \frac{\alpha}{30\pi}\omega_l\sin^2\frac{\vartheta}{2}\sin 2\sigma \int\!\ud x^\LCp\ \frac{E^2(x^\LCp)}{E_S^2} \;.
\ee
We have a straightforward factorisation of dependencies, and the probability is clearly maximal for head-on collisions, $\vartheta=\pi$, and a $45^\circ$ angle between the probe polarisation and the background, $\sigma=\pi/4$; this is well known for birefringence. This leaves us with the pulse shape. For the above angles, consider first a {\it constant} field of duration $\Delta x^\LCp$ in lightfront time. Introducing the probe wavelength $\lambda_l = 2\pi/\omega_l$, we find
\be\label{epe}
	(\ref{S})  \to \frac{\alpha}{15}\frac{\Delta x^\LCp}{\lambda_l}\bigg(\frac{E}{E_S}\bigg)^2 = \frac{2\alpha}{15}\frac{d}{\lambda_l}\bigg(\frac{E}{E_S}\bigg)^2\;,
\ee
where, in the last step, we have introduced the distance $d$ traveled by the probe in the ``birefringent medium''; for the case of a head-on collision between a photon and a plane wave, this distance is $d=\Delta x^\LCp/2$, as shown in Fig.~\ref{rumtid}.  We recognise the structure of the birefringence signal~\cite{Heinzl:2006xc}, which has come directly from the flip amplitude in a background field. Note that the observable is (\ref{epe})-squared. In the quantum theory, this is the probability of polarisation-flip. In the effective theory it is the intensity of the induced component of the probe beam, which is proportional to the phase retardation squared~\cite{Heinzl:2006xc}.

It is clear that the constant field probability can be increased by raising the field's amplitude  or extending its duration, so let us now consider a nonconstant wave of arbitrary profile. Returning to (\ref{S}), we could calculate the $x^\LCp$--integral for different intensities and field configurations and then compare. However, the physically realistic scenario is that a laser will have a fixed amount of energy which can be manipulated (e.g.\ via focussing) into different pulse shapes. For phenomenological comparisons between pulses, the parameter we should therefore keep constant is the energy. The best we can do in a plane wave is to hold constant the total energy per transverse area (or, integrated intensity), which is
\be\label{energi}
	\frac{1}{2}\int\!\ud x^\LCp [E^2(x^\LCp) + B^2(x^\LCp)] = \int\!\ud x^\LCp E^2(x^\LCp) \;.
\ee
We see immediately that if the energy is fixed, (\ref{S}) is also fixed, and the flip probability becomes independent of the pulse shape, under the constraint of fixed energy. This is a {\it potentially} positive result, as it suggests that birefringence signals, while weak, are robust. The important question is of course how well this result extends to more realistic, focussed background fields. This will be investigated elsewhere.

\begin{figure}[t!]
	\includegraphics[width=\columnwidth]{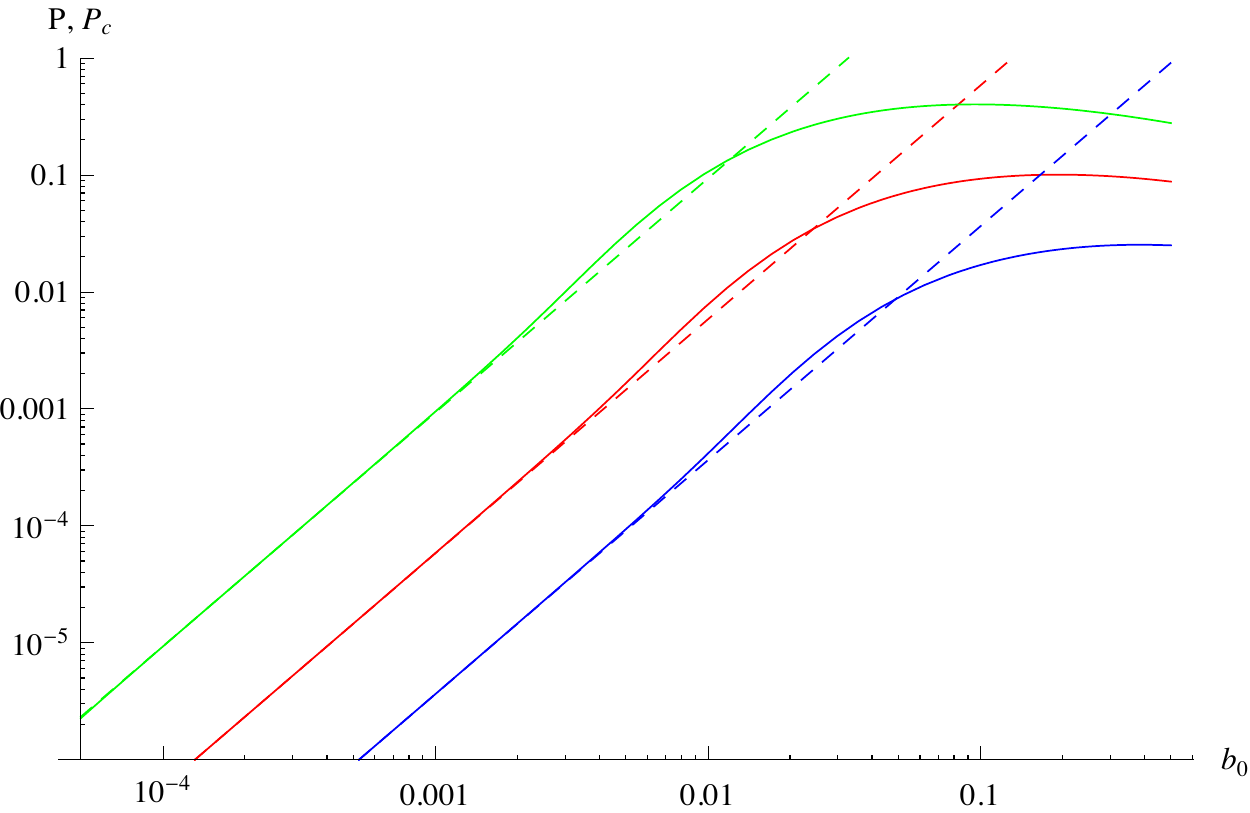}
		\caption{\label{FIG:P-vs-P} The exact probability for polarisation flip $\epsilon\to \epsilon'$ (with $\epsilon'\epsilon=0$), from (\ref{PFLIP}), and the low energy approximation (\ref{momentum expansion}), as a function of invariant $b_0=kl/m^2$ (log scale). The background has linear polarisation, a Gaussian profile with width $\Lambda=10\pi$ and $a_0=200,100, 50$ (left to right).}
\end{figure}

\begin{figure}[ht!]
		\includegraphics[width=\columnwidth]{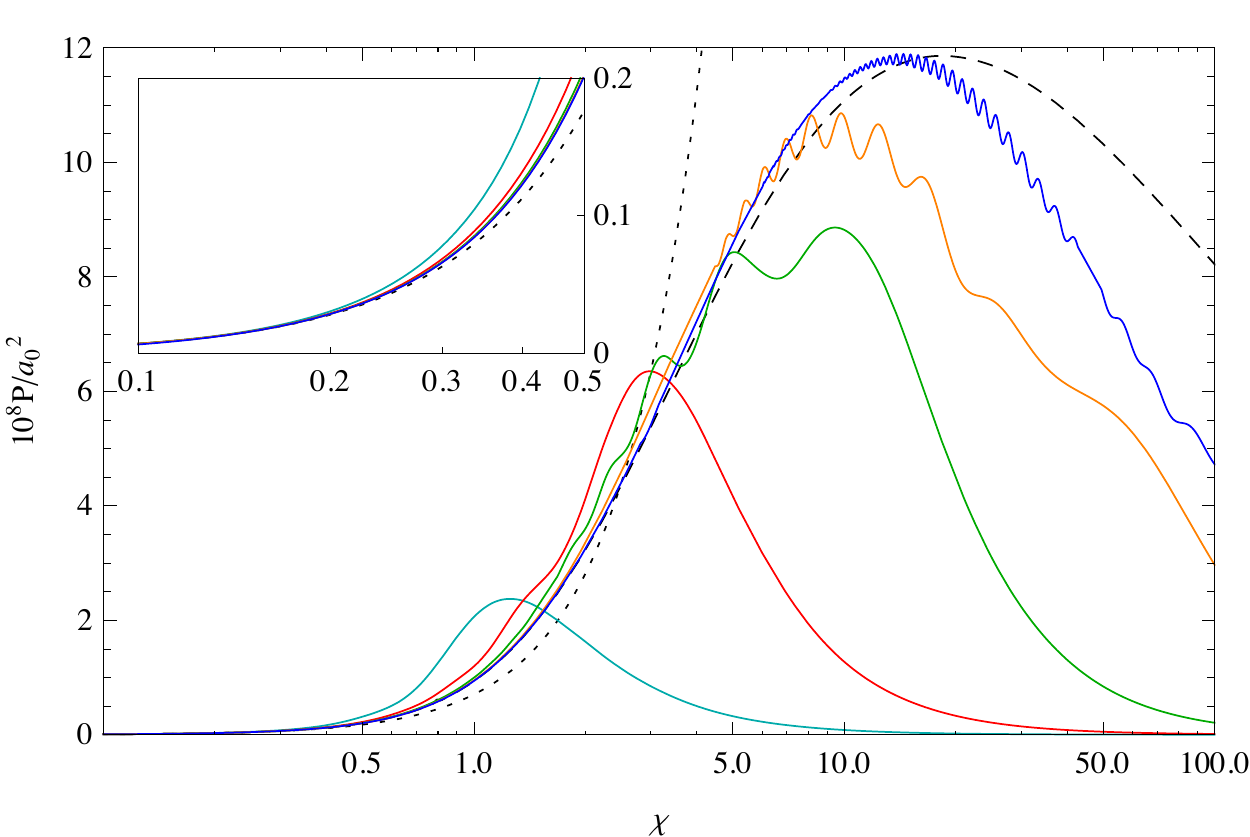}
		\caption{\label{FIG:SCALING} The exact probability for polarisation flip $\epsilon\to \epsilon'$ (with $\epsilon'\epsilon=0$), as a function of $\chi$, for $a_0 = \tfrac{1}{2}, 1, 2, 4, 8$. The dotted line is the low energy approximation and represents low energy scaling, see inset. The black dashed line is $a_0=1000$, demonstrating the crossed field, or large $a_0$, limit with high intensity scaling.}
\end{figure}

The probability for polarisation flip from (\ref{PFLIP}) and its approximation (\ref{momentum expansion}) are compared in Fig.~\ref{FIG:P-vs-P}. We take here a linearly polarised background given by (\ref{profile}) with $\varphi=\pi/4$ and both $\phi_{0i}=0$. Peak $\chi=a_0 b_0$ for linear polarisation. The approximation (\ref{momentum expansion}) is within $10\%$ of the full probability for $\chi<0.6$, and the worst underestimate, (\ref{momentum expansion})/(\ref{PFLIP})$\sim0.74$, occurs at $\chi\simeq 1.1$. We saw above that, in the low energy regime, only the total pulse energy is relevant, and this is proportional to the total, integrated $\chi$. Hence there is a scaling behaviour in the low energy regime: the ratio $\mathbb{P}_\text{flip}/a_0^2$ depends only on the product $\chi = a_0 b_0$. This behaviour extends to {\it all} energies in the case of crossed fields as can be seen explicitly in e.g.\ the refractive indices for constant fields~\cite{Toll:1952,Narozhny:1968,Ritus:1972,Heinzl:2006xc,Shore:2007um}. In more realistic pulsed fields, though, there are scaling violations at high energies, essentially because the space-time variations of the background induce $b0$-dependent corrections to constant field results.  In Fig.~\ref{FIG:SCALING} we have plotted the flip probability as a function of peak $\chi=a_0 b_0$. Polarisations are as above. We have chosen a selection of lower  intensities $a_0=\mathcal{O}(1)$ in order to show the structure of the probability. The scaling behaviour is clear for low $\chi$, where all the curves sit on top of each other (see the inset). At high energy we see scaling violation, as the probabilities become dependent on field-strength ($a0$) and probe energy ($b_0$) separately. The probability exhibits a great deal of structure when the energy scale dominates over the intensity scale, i.e.\ when $b_0 > a_0$. More about this structure can be found below and in Appendix~\ref{APP-C}. These features are washed out in the large $a_0$ limit in which the intensity once again dominates, and the crossed field result is expected to be a good approximation~\cite{Nikishov:1963,Nikishov:1964a}. This is corroborated by our results for the probability when $a_0=1000$ (black, dashed line).

\section{Forward scattering, the refractive indices and pair production.}\label{SECT:NON}
We discuss here the non-flip forward scattering amplitude, and its relation to pair production and the refractive indices, again for arbitrary plane waves.  The non-flip amplitude is of theoretical interest because it gives, as described in Sect~\ref{SECT:NEW}, the spacetime-dependent refractive indices, and because together with the flip amplitude it completely defines the polarisation tensor for on-shell external photons via (\ref{TENSOR}).

For a one-photon state $\ket{f,\epsilon}$ of form (\ref{foton-in}), the one-loop forward scattering amplitude is
\be\begin{split}
	\mathcal{F} &:= \bra{f,\epsilon} S\ket{f,\epsilon} \\
	&= 1 + i\int\!\ud{\sf l}\ |f(l)|^2\ \bar{\epsilon}\,\Pi(l,l)\epsilon\;.
\end{split}
\ee
There are two (independent) such amplitudes, associated with the photon's two (independent) polarisation/helicity states. The relevant Feynman diagram is as in Fig.~\ref{FIG:LOOP}, but with the same incoming and outgoing states.  Taking the wavepacket to be peaked around $l_\mu$, the non-flip forward scattering amplitude is, in terms of $T$,
\be\label{F1-resultat}
	\mathcal{F} = 1 + i T_{\epsilon\epsilon}(\infty) \;.
\ee
In this case, one can have contributions from all the terms in (\ref{T-resultat}). Our results in this section can be verified by comparison with results for limiting cases derived previously, e.g.\ the constant crossed field case~\cite{Narozhny:1968,Shore:2007um} (see Appendix~\ref{APP:CROSSED}).

\subsection{The refractive indices}
For an incoming monochromatic probe with momentum $l_\mu$ and polarisation $\epsilon_\mu$, the refractive indices are, from (\ref{F1-n}),
\be\label{F1-n2}
	\mathrm{n}(x^\LCp) = \frac{|{\underline\wavevector}|}{\wavevector_0}  = 1 +\frac{nl}{\omega_l^2}\frac{\partial T_{\epsilon\epsilon}(x^\LCp)}{\partial x^\LCp} \;.
\ee
Inserting the low energy approximation (\ref{Low energy R}) into this formula gives us the following expressions for the refractive indices. First, if we again take $\vartheta$ to be the angle between the directions of the probe and the background momentum (three) vectors, and $\sigma$ to be the angle between the polarisation vectors of the probe and the background, then
\be
	n(x^\LCp)=1+\frac{2\alpha}{45\pi}\sin^2\frac{\vartheta}{2}(7-3\cos^2\sigma)\frac{E^2(x^\LCp)}{E_S^2} \;.
\ee
In particular, if we take $\epsilon a=0$ then we find
\be\begin{split}\label{n-perp}
	 \mathrm{n}_\perp(x^\LCp)  &= 1 + 7\frac{\alpha}{90\pi}\frac{m^2\chi^2(x^\LCp)}{\omega_l^2} \;,
\end{split}
\ee
while for $\epsilon \parallel a$ we have
\be\begin{split}\label{n-par}
	 \mathrm{n}_\parallel(x^\LCp)  &= 1 + 2\frac{\alpha}{45\pi}\frac{m^2\chi^2(x^\LCp)}{\omega_l^2} \;,
\end{split}
\ee
The low-energy indices are independent of the probe frequency $\omega_l$ (as it cancels against $\chi$, see (\ref{F1-n2})), but are dependent on pulse shape and position $x^+$ (see also~\cite{Gies:2011he}).  For the case of a constant field the indices become constant and recover (28) in~\cite{Narozhny:1968}. Note that it is $T_{\epsilon\epsilon}$ which appears in asymptotic observables; this ``integrated index'' is just the difference in optical path lengths in the birefringent vacuum, compared with the ordinary vacuum:
\be \label{optPL}
	\frac{nl}{\omega_l^2} \, T_{\epsilon\epsilon}(\infty) = \int\! \ud x^\LCp\ (\mathrm{n}-1)
\ee
For low energy, it is proportional to the energy seen by the probe. We therefore find again that, in the limit that $kl/m\ll 1$, the relevant property of the pulse is its total energy.
\subsection{Behaviour as a function of energy}
The imaginary part of the non-flip amplitude obeys the well-known optical theorem result that, for all energies,
\be\label{IM-PAR}
	2\, \text{Im } T_{\epsilon\epsilon}(\infty) = \mathbb{P}_\epsilon( \text{pair} ) \;, 
\ee
where $\mathbb{P}_\epsilon(\text{pair})$ is the total probability of stimulated pair production by a photon of polarisation $\epsilon$~\cite{Nikishov:1963,Nikishov:1964a,Schutzhold:2008pz,Heinzl:2010vg,Nousch:2012xe,Titov:2012rd}, written here using the expression (\ref{T-resultat}), not previously found in the literature, in which the final state momentum integrals have been performed. The small $b_0$ expansion of $\text{Im } T_{\epsilon\epsilon}(\infty)$ vanishes, because pair production is nonperturbative in the kinematic invariant $b_0$. This can be seen as follows. In vacuum, pair production by two photons with momenta $k_\mu$ and $l_\mu$ is a threshold process forbidden for $kl<2m^2 \iff b_0<2$. For a photon $l_\mu$ in a background, this threshold is removed because of the (in principle) arbitrarily high frequency components of the background; if $k_\mu$ is here the {\it central} frequency and $r$ parameterises the frequency range, the $S$-matrix element will contain the structure
\be
	\int\!\ud r\ \theta_\text{step}\big( r kl/m^2-2\big)\cdots 
\ee
While this integral does have support for arbitrarily low probe energy $\omega_l$, it remains identically zero to all orders in a low energy expansion.

We could plot the refractive indices as a function of probe energy and lightfront time, but since it is the optical path length $T_{\epsilon\epsilon}(\infty)$ which enters asymptotic observables, recall (\ref{optPL}), we focus on this. We will therefore examine the real and imaginary parts of the forward scattering amplitude (\ref{F1-resultat}) as a function of probe energy. This requires a  careful and sophisticated numerical evaluation of (\ref{T-resultat}), see Appendix~\ref{APP-C}, and in preparation it is worth comparing the large-$\phi$ behaviour of the integrand in (\ref{T-resultat}) for small and large $b_0$. Outside of the volume for which the pulse is peaked, the general behaviour of the integrand is to decrease with increasing $\phi$. At low $b_0$, the decrease is rapid, occurring in a region not much larger than that in which the pulse is peaked. In the ``fully quantum regime'' of large $b_0\gg 1$, though, the integrand decreases more slowly, with the $\theta$-integral contributing many oscillations, and the effect of the pulse is noticeable in the integral even far outside the pulse volume. Consequently, the integral receives large contributions from coherent/interference effects at large $b_0$. (Compare large $a_0$, in which the locally constant, or incoherent, approximation applies.). This behaviour is also seen in the effective mass $M^2$, which is nonzero even when one of its arguments is far outside the pulse~\cite{Kibble:1975vz, Hebenstreit:2010cc}.

We will consider a linearly polarised probe and three different circularly polarised field profiles. The first is monochromatic; set $\Lambda=\infty$ in (\ref{profile}), and the others are pulsed, with $\Lambda=10\pi$ and $\Lambda=\pi$, the latter being a very short pulse. (We remark that carrier phase effects may become particularly important in short pulses, although we do not analyse this here. Further, for more on the impact of photon polarisation, see e.g.~\cite{King:2013zw}.)

\begin{figure}[t!]
\includegraphics[width=\columnwidth]{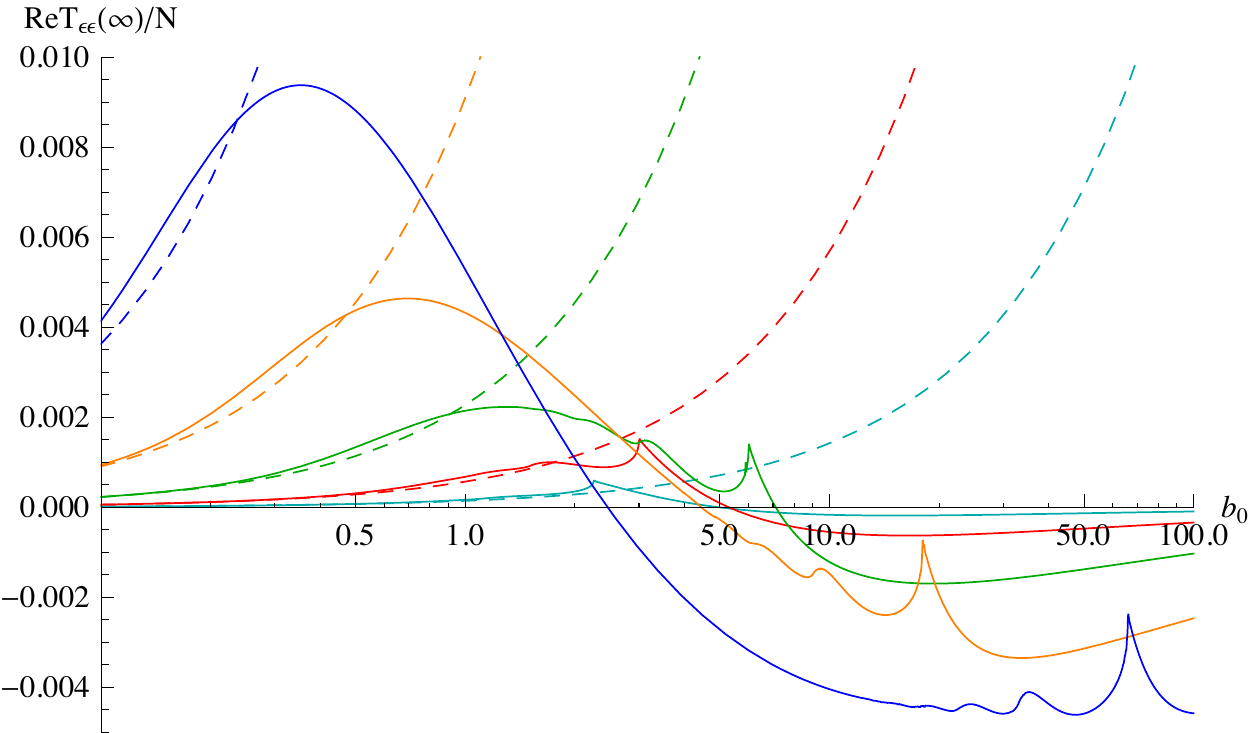}
\includegraphics[width=\columnwidth]{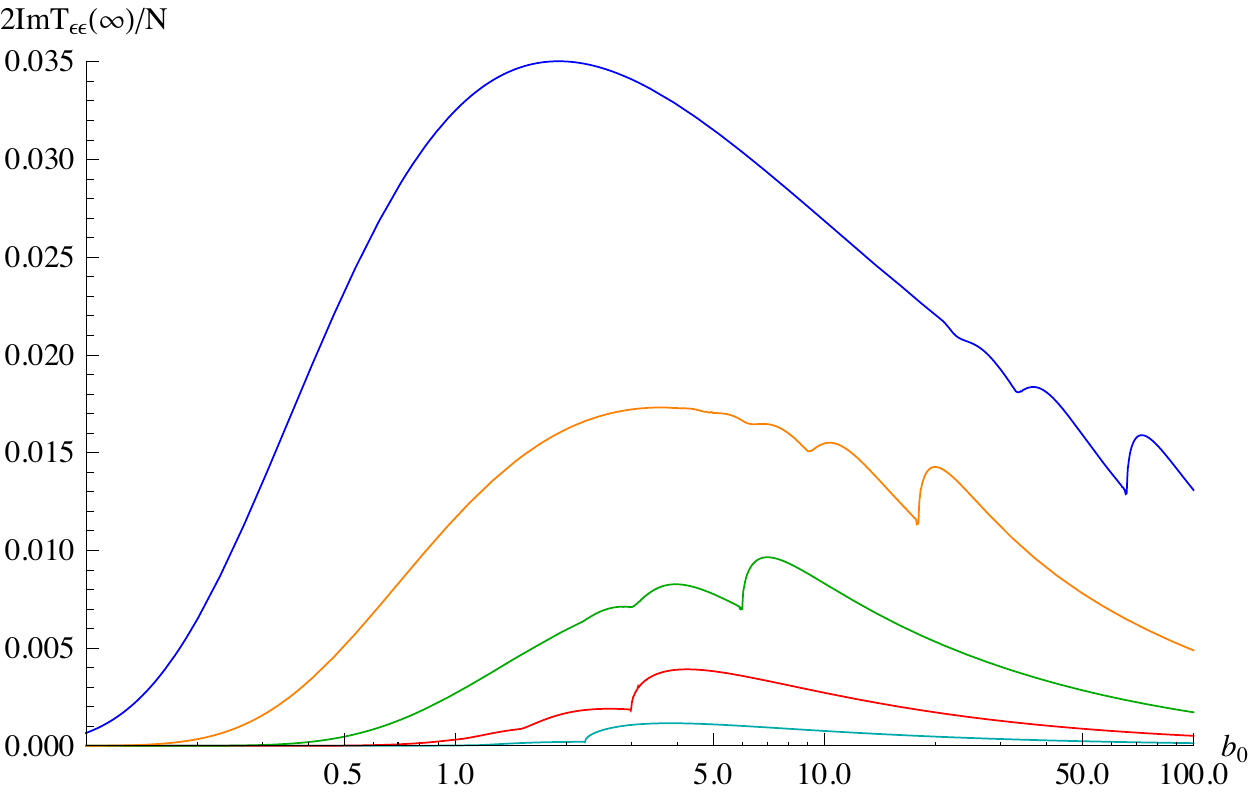}
\caption{\label{FIG:MONO} The real and imaginary parts of the forward scattering amplitude for a circularly polarised, monochromatic wave, (\ref{profile}) with $\Lambda=\infty$, $a_0=1/2$ (cyan), 1 (red), 2 (green), 4 (orange), 8 (blue).}
\end{figure}

For monochromatic backgrounds the integrand in (\ref{T-resultat}) depends periodically on $\phi$, so finite quantities can be obtained only per cycle, i.e.\ by restricting the $\phi$--integral to one cycle. The real and imaginary parts of $T_{\epsilon\epsilon}(\infty)/N$, for $N$ the total number of cycles, are plotted in Fig.~\ref{FIG:MONO}. Before considering their detailed structure, we begin with the broad behaviour of the amplitude (which is common to all three background profiles). The imaginary part (the probability per cycle of pair production) is positive, initially small, and then rises steeply. The real part (the integrated real part of the refractive index, or optical path difference) begins positive (normal dispersion), peaks, and then decreases (anomalous dispersion) as the pair production probability reaches its maximum. For higher energies the real part becomes negative, corresponding to $\text{Re }\mathrm{n}<1$. This behaviour mirrors that of the refractive indices in a constant field, and is typical of a medium with a single absorption band as described in~\cite{Shore:2007um}. 

We have though additional structure in comparison to the constant field case. For a monochromatic field, one expects threshold behaviour in the pair production probability, corresponding to resonance conditions~\cite{Heinzl:2010vg} which describe the effective conservation of quasi-momentum~\cite{Nikishov:1963,Nikishov:1964a},
\be\label{quasi}
	q_\mu + q'_\mu = l_\mu + n k_\mu \;, \quad n = 1,2,3\ldots \;,
\ee
where the quasimomenta obey $q^2=m_*^2$, the effective mass is $m_*^2 = m^2(1+a_0^2/2)$~\cite{Sengupta,Volkov,Kibble:1965zz}, see (\ref{profile}),  and $n k_\mu$ represents discrete energies taken from the background. Squaring (\ref{quasi}), this implies the threshold values 
\be\label{threshold}
	b_0 = \frac{2m_*^2}{n} \;,
\ee
which are clearly visible in the pair production probability, bottom panel of Fig.~\ref{FIG:MONO} for low $n=1,2,3$, with the peak positions moving from right to left with increasing $n$, at fixed $a_0$. For higher $n$, i.e.\ further to the left, the peaks are washed out by many overlapping contributions, a behaviour typical also of nonlinear Compton scattering in a monochromatic wave~\cite{Harvey:2009ry}. While the $n=1$ threshold is sometimes associated with the minimum energy needed to create a pair, we clearly see that higher order effects can sum up to give comparable contributions at much lower probe energies. There is a corresponding behaviour in the real part of the forward scattering amplitude~\cite{Becker-Mitter}. The real part peaks sharply upwards as the imaginary part dips, which happens just before an effective mass threshold is reached. So, when the absorptive part dips, (\ref{quasi}) and (\ref{threshold}) lead to a resonance peak in the dispersive part.  Interestingly, for larger laser intensity $a_0$, the peaks (dips) in the real (imaginary) part become more pronounced, and the forward scattering amplitude begins to resemble the refractive index in a medium with double band structure, in which one can have `transparency'. This is the vanishing of the absorptive part $\text{Im }\mathrm{n}=0$, which would correspond here to a vanishing pair production probability. See~\cite{Shore:2007um} for a discussion of scenarios in which this could be realised.

\begin{figure}[t!]
\includegraphics[width=\columnwidth]{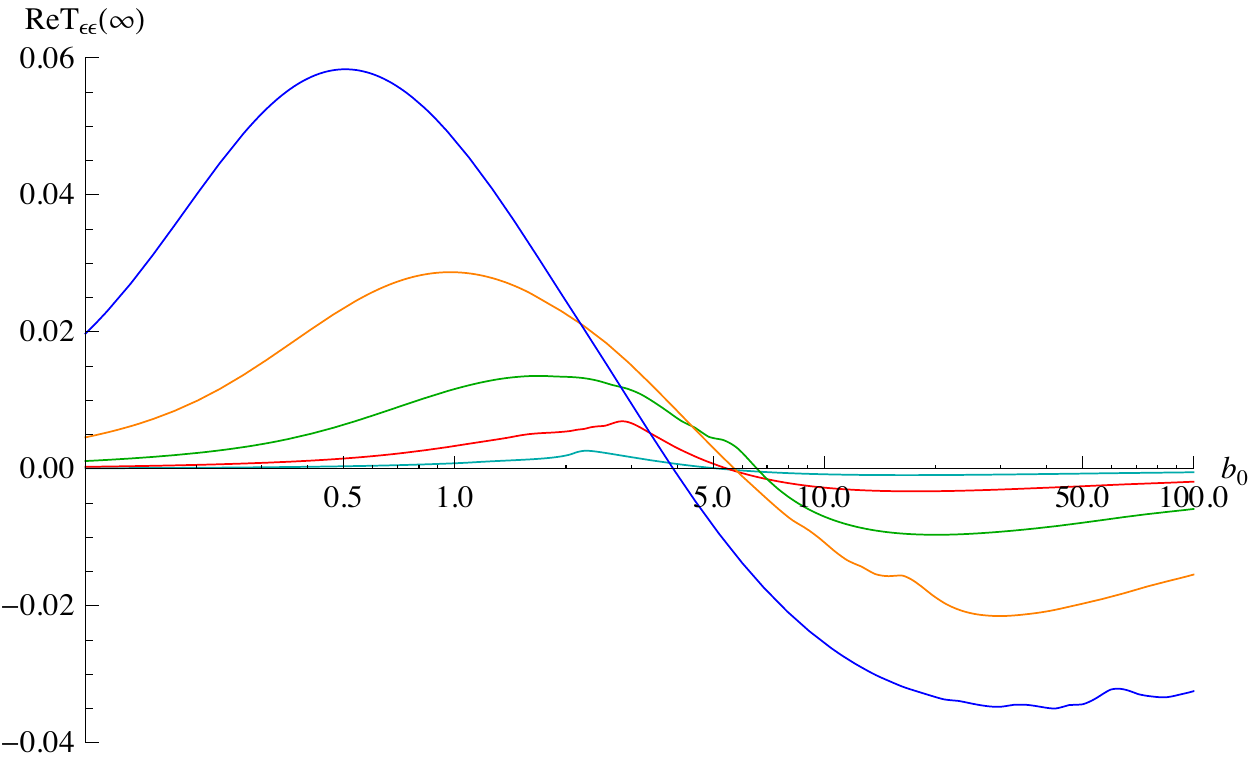}
\includegraphics[width=\columnwidth]{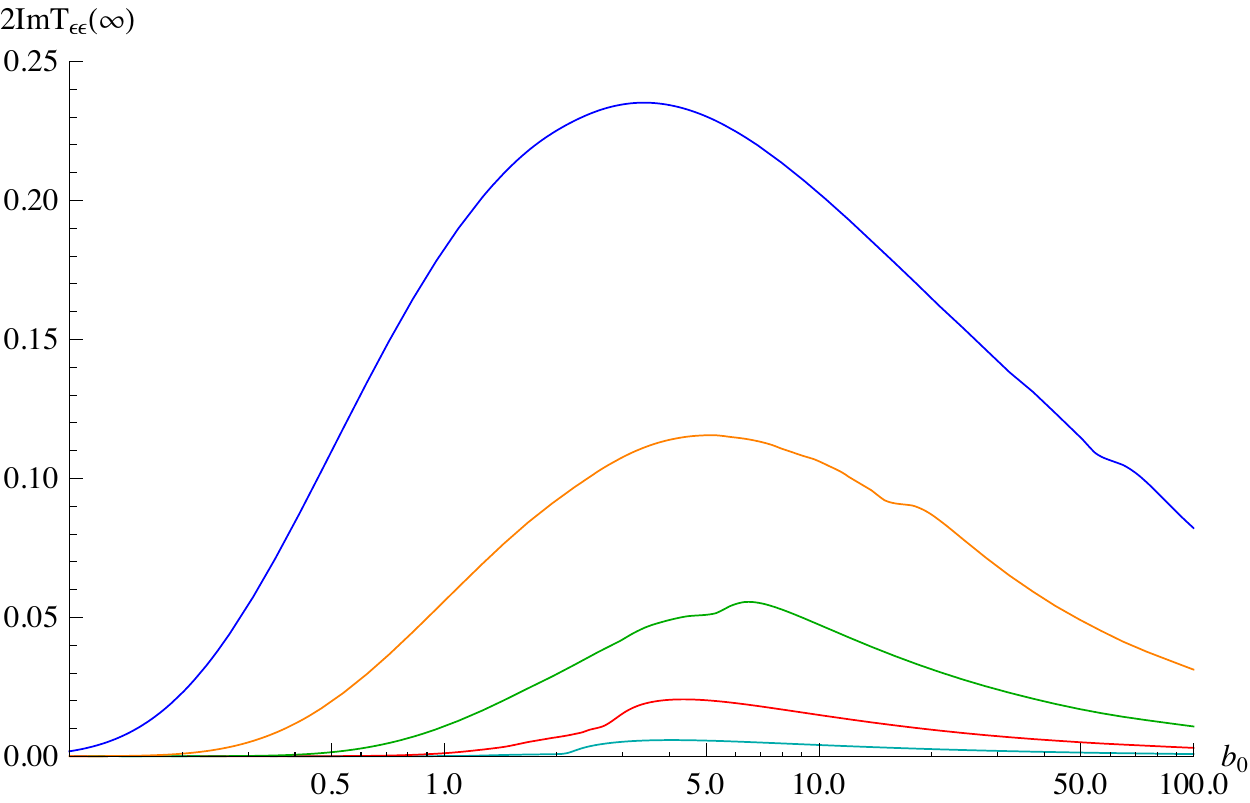}
\caption{\label{FIG:LONG} The real and imaginary parts of the forward scattering amplitude for a circularly polarised, Gaussian pulse (\ref{profile}) with $\Lambda=10\pi$, $a_0=1/2$ (cyan), 1 (red), 2 (green), 4 (orange), 8 (blue).}
\end{figure}

\begin{figure}[t!]
\includegraphics[width=\columnwidth]{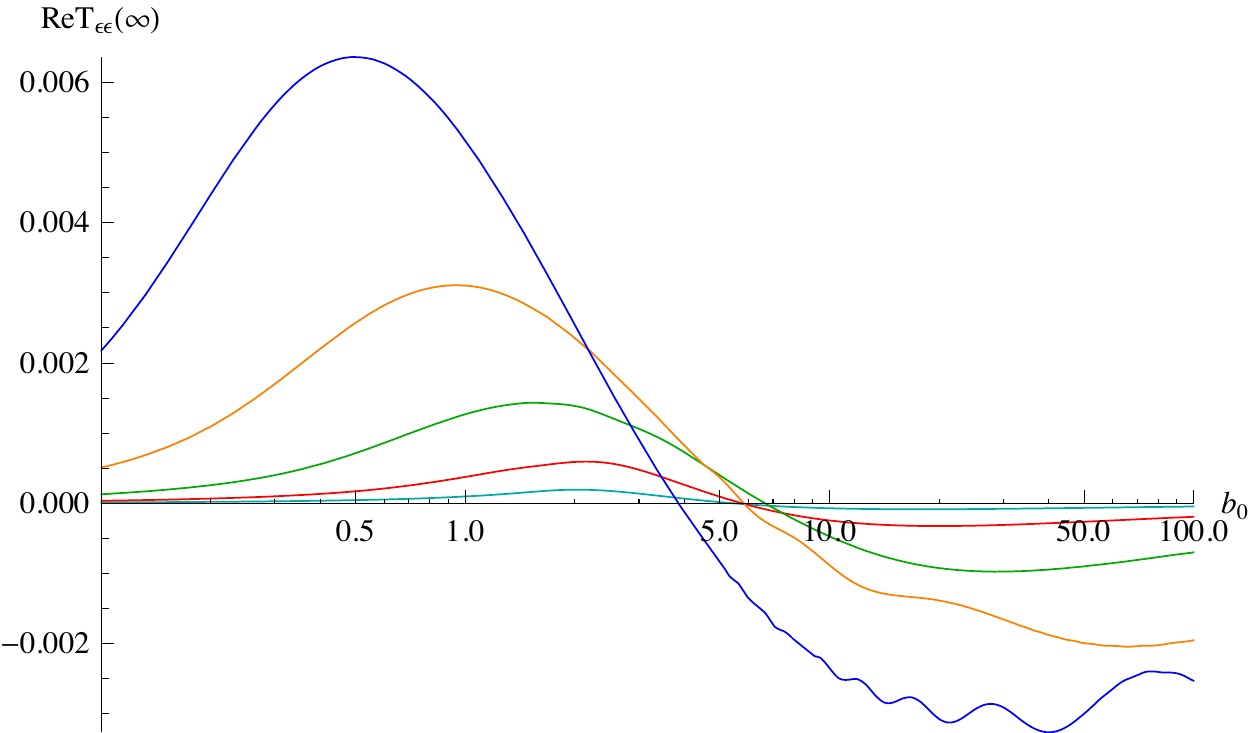}
\includegraphics[width=\columnwidth]{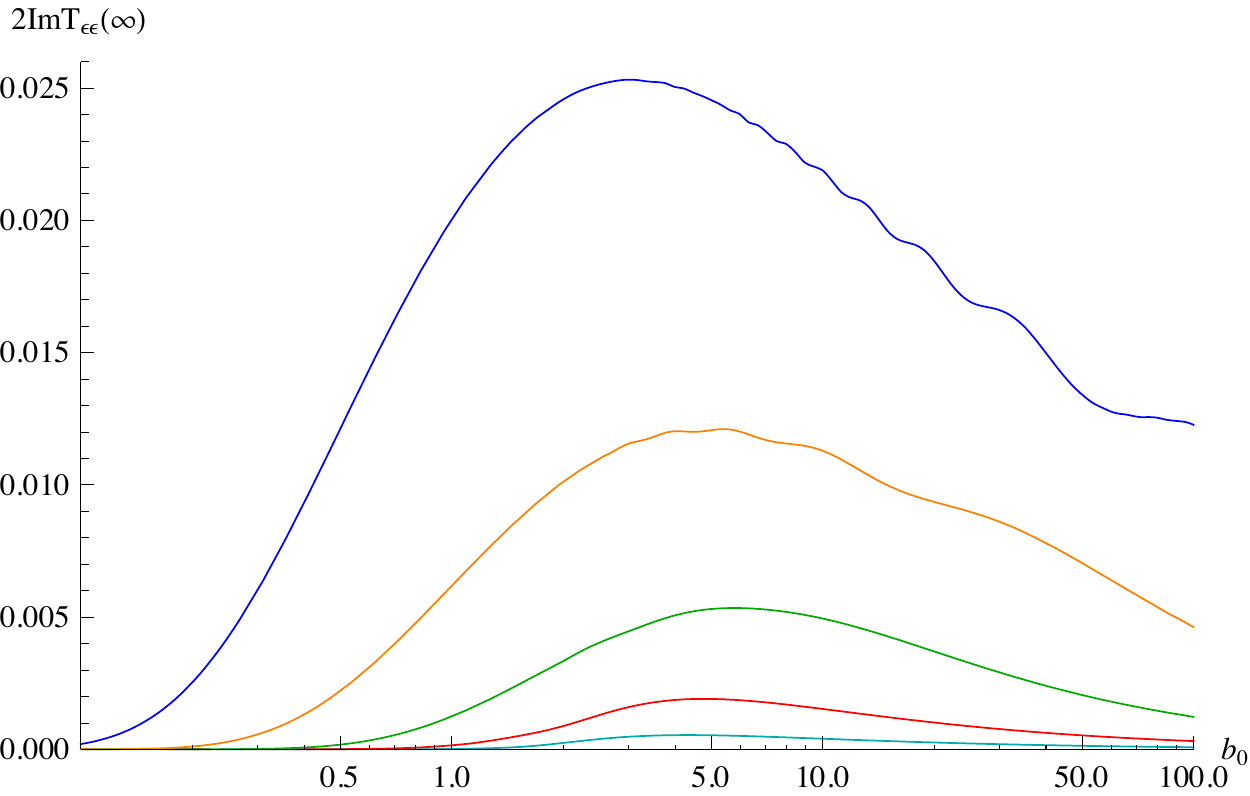}
\caption{\label{FIG:SHORT} The real and imaginary parts of the forward scattering amplitude for a circularly polarised, short Gaussian pulse (\ref{profile}) with $\Lambda=\pi$, $a_0=1/2$ (cyan), 1 (red), 2 (green), 4 (orange), 8 (blue).}
\end{figure}

We now go to pulsed fields, beginning with $\Lambda=10\pi$. The forward scattering amplitude $T_{\epsilon\epsilon}(\infty)$ is plotted in Fig.~\ref{FIG:LONG}. In this case, we see that the broad behaviour of the forward scattering amplitude is as before. However, structure due to the effective mass has largely been washed out, because we lose periodicity by going to pulses with quickly-dropping envelope functions such as Gaussians~\cite{Harvey:2012ie}. It is possible though to retain such structure, if one wishes, by tailoring the pulse shape.  A longer pulse with a reasonably flat envelope, which contains many similar oscillations, will retain some features of the monochromatic case, interpolating between the monochromatic and pulsed results in Fig.'s~\ref{FIG:MONO} and \ref{FIG:LONG}. For explicit examples see~\cite{Harvey:2012ie} for nonlinear Compton scattering, \cite{Heinzl:2010vg} for stimulated pair production and \cite{Kohlfurst:2013ura} for Schwinger pair production.

The forward scattering amplitude in the short pulse with $\Lambda=\pi$ is plotted in Fig.~\ref{FIG:SHORT}. In this short pulse, a new structure arises. This can be seen in e.g.\ the $a_0=8$ curve. Unlike the abrupt cusps signalling the effective mass, there is in the high-energy regime, $b_0\gg1$, an oscillation which builds smoothly in amplitude. This can likely be attributed to an increased importance of high--frequency components in the background, the interaction of many such components with the probe giving rise to the structure shown. We find in particular that, at large $b_0$, the most significant contributions to the integrals come from the regions when $\phi\pm\theta/2$ (which are the original, ordered times appearing in the $S$-matrix element) lie outside and on opposite sides of the pulse peak. One might say that this occurs when the virtual particles running the loop see the whole pulse before annihilating. This should be contrasted to the low energy, but high $a_0$ behaviour in~\cite{DiPiazza:2013iwa}; in that case the virtual pair typically annihilates after a short time.

\begin{figure}[t!]
\includegraphics[width=\columnwidth]{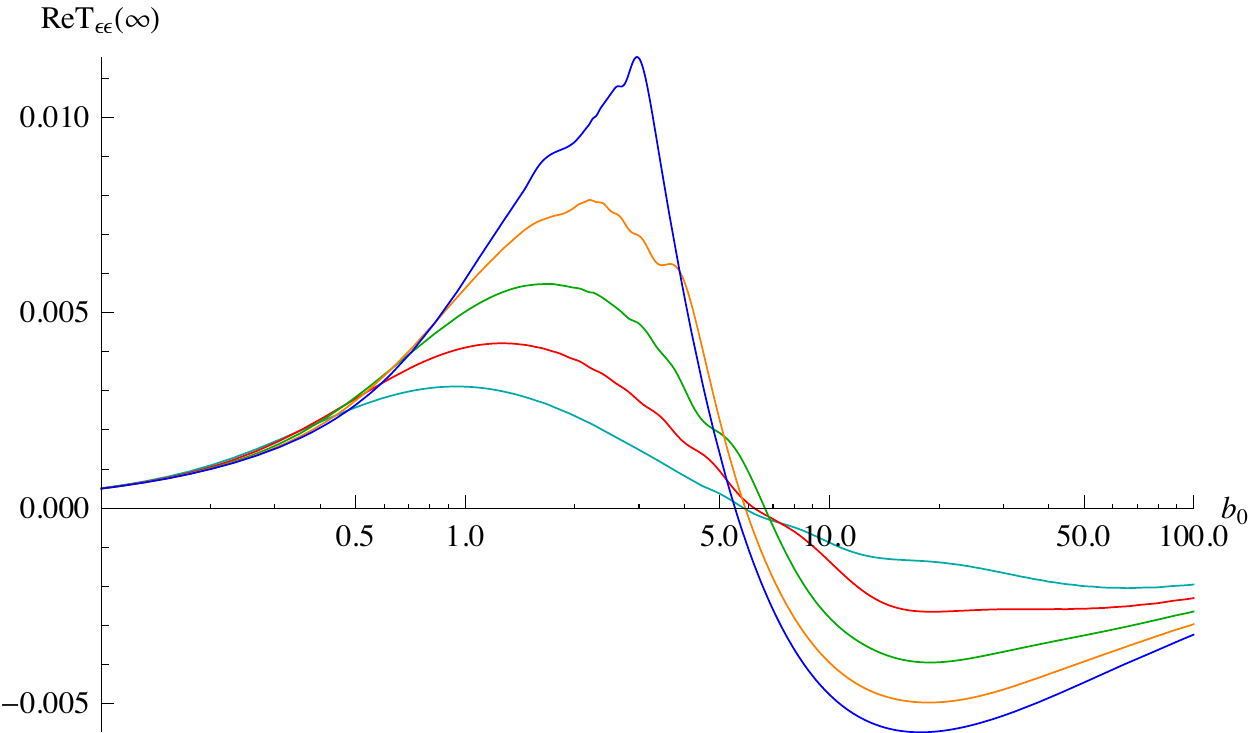}
\includegraphics[width=\columnwidth]{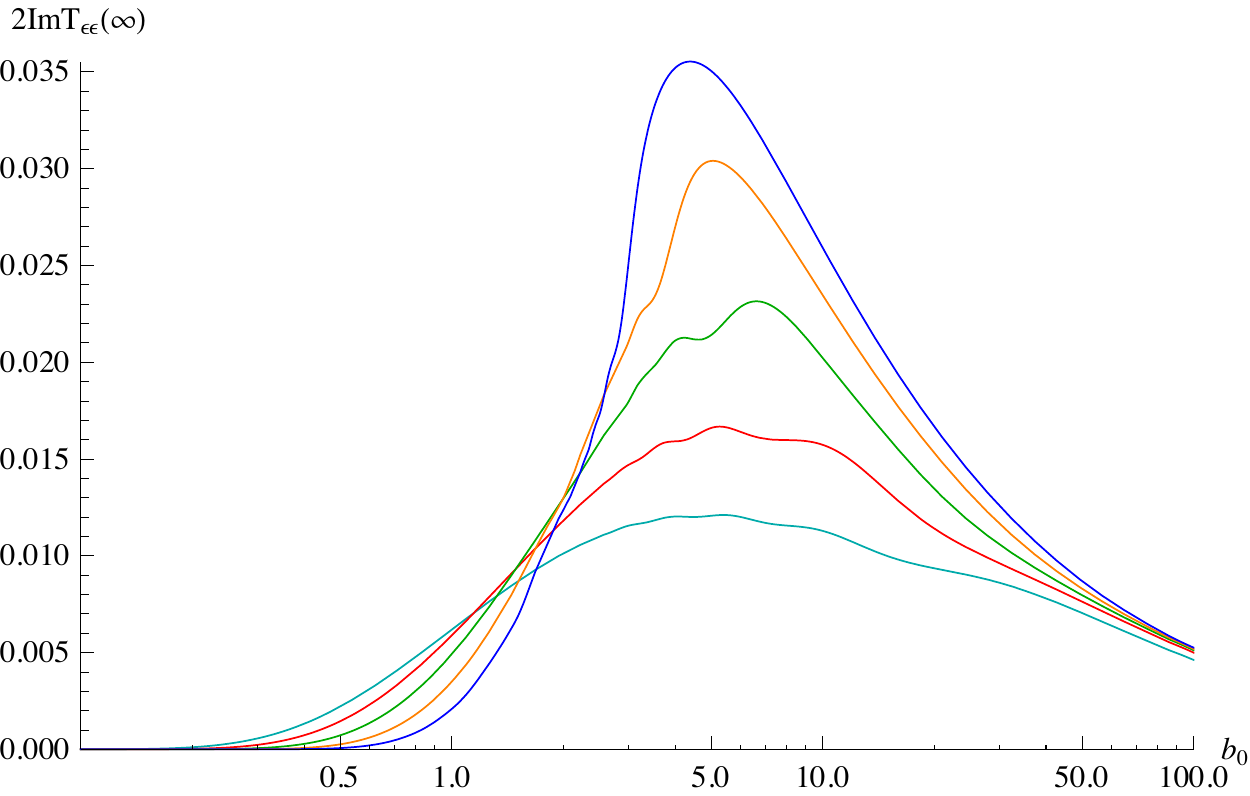}
\caption{\label{FIG:ENERGI} The real and imaginary parts of the forward scattering amplitude for a circularly polarised pulse with constant energy, shaped into different profiles with $(a_0,\Lambda/\pi) = $ (4,1) (cyan), (2.93, 2), (2.09,4), (1.48,8), (1.05,16)  (blue).}
\end{figure}

Finally, we study the effects of pulse shaping under the reasonable physical assumption that the total pulse energy (\ref{energi}) is constant. We plot in Fig.~\ref{FIG:ENERGI} the forward scattering amplitude for various intensities $a_0$ and pulse lengths $\Lambda$ such that the total pulse energy is constant. The scaling law for low energy probes is clearly visible in the real part; for fixed energy, the results are independent of pulse shape. This scaling law is violated at high energies. The behaviour of the imaginary part gives the following insight into realisations of stimulated pair production. As a function of probe energy, the probability of pair production in the high intensity/short pulse (cyan curve) increases most rapidly from zero. As probe energy increases, though, the low intensity/long pulse probability rises to much greater values. Hence, if available probe energy is low, it is better to use a short intense pulse. If probe energy is higher, it is better to use a long, less intense pulse.

\section{Conclusions}
We have presented the polarisation-flip and non-flip amplitudes for a photon of arbitrary momentum $l_\mu$ probing an arbitrary plane wave background with central momentum $k$. We have obtained expressions which are simple and compact. All these amplitudes exhibit a rich structure for moderate intensities and high energies, and contain parts which are nonperturbatively small in the kinematic invariant $kl/m^2$. All the amplitudes admit simple and accurate low-energy approximations in the (low energy) kinematic regime of relevance to upcoming birefringence experiments. This confirms the validity of previous analyses based on calculating directly in the low-energy limit.

From the non-flip forward-scattering amplitudes we have extracted the effective refractive indices of the vacuum, going beyond previous constant field results. The behaviour of the non-flip amplitude, as a function of probe energy, reflects that of the refractive indices, exhibiting first normal, then anomalous, dispersion, together with absorption due to pair production, as well as (depending on pulse shape) threshold and resonance effects due to the intensity-dependent effective mass.
 
We have concentrated on polarisation effects without changes in momentum. As we saw, a single photon is always scattered forward in a plane wave background. In a general background, though, it may be that forward scattering is more likely than non-forward scattering, or vice versa, and this may provide clues for identifying the most promising experiment to detect effects due to light-by-light scattering. Alternatively, it is easy to imagine that the field structure could be tuned to emphasise a chosen effect. These are interesting questions for the future.

\appendix

\section{Strong field QED in lightfront quantisation}\label{APP-A}
The most convenient formalism with which to study physics in plane wave backgrounds is, as first advocated in~\cite{Neville:1971uc}, lightfront field theory, in which one quantises on lightlike hyperplanes, taking $x^\LCp= x^0+x^3$ as the time direction. (See \cite{Brodsky:1997de,Heinzl:1998kz,Heinzl:2000ht} for comprehensive reviews.) Coordinates are $x^{\LCpm}=x^0\pm x^3$, $x^\LCperp=\{x^1,x^2\}$, $p_\LCpm=(p_0\pm p_3)/2$, $p_\LCperp=\{p_1,p_2\}$.

There are several advantages to using this `front form' over the usual `instant form' of quantisation. First, we can choose coordinates so that the background's phase is aligned with (it depends only on) $x^\LCp$, which makes calculations simple because the background enters only through time-dependent factors. Second, because lightfront field theory is an on-shell formalism, all of the $p_\LCp$ integrals in Feynman diagrams are done for free. Third, many terms vanish immediately because $p_\LCm>0$, which is related to the properties of the ligthfront vacuum.

One disadvantage of the formalism is that a Feynman diagram with $n$ vertices leads to (roughly) $n!$ lightfront-time-ordered diagrams. Even though this increase is minimal for the low-order processes we consider here, it is nevertheless simplest to drop the diagrammatic approach entirely, and calculate directly with the states and Hamiltonian. This may be viewed as a particular incarnation of `old-fashioned perturbation theory' \cite{Weinberg:1995mt}. We give now a brief overview of the theory, in preparation for the calculation below.

The mode expansion for the gauge field is
\be
	A^\mu = \int\!\ud{\sf l}\ a_\lambda(l)\epsilon^\mu_\lambda(l) e^{-ilx} + \text{ c.c.,}
\ee
with  $A^\LCp =0$ (lightfront gauge), $A^\LCm = \partial_\LCperp A_\LCperp/\partial_\LCm$ the constrained field, and $\epsilon_\lambda$  the helicity state vectors. The invariant, on-shell momentum space measure $\ud{\sf p}$ is, in lightfront coordinates,
\be
	\int\!\ud{\sf p} \equiv \int\!\frac{\ud^2p_\LCperp}{(2\pi)^2}\int\limits_0^{\infty}\!\frac{\ud p_\LCm}{(2\pi)2p_\LCm} \;,
\ee
and $p_\LCp$ is always given by the mass-shell condition, $p_\LCp = (p_\LCperp^2+m^2)/(4p_\LCm)$, $m\to 0$ for photons. We define $a^\mu(l) = a_\lambda(l)\epsilon^\mu_\lambda(l)$. The commutator of the photon modes is
\be
	[a_\mu(l),a^\dagger_\nu(l')] =  -\delta({\sf l},{\sf l}')\bigg(g_{\mu\nu} - \frac{n_\mu l_\nu + l_\mu n_\nu}{nl} \bigg) \;,
\ee
where $\delta({\sf l}',{\sf l})$ is the relativistically invariant delta function,
\be
	\int\!\ud{\sf l}\ \delta({\sf l},{\sf l'}) = 1 \;,
\ee
which, in lightfront coordinates, has the form
\be
	\delta({\sf l}',{\sf l}) = 2l_\LCm (2\pi)^3\delta^2(l'_\LCperp-l_\LCperp)\delta(l'_\LCm-l_\LCm)  \;.
\ee
For the fermion fields we use the four-component, rather than two-component, spinor formalism,
\be
	\Psi(x) = \tint{p}\ b_\lambda(p) \varphi_{p} K_p u_{\lambda p} + \bar{K}_p v_{\lambda p} d_\lambda^\dagger(p) \varphi_{-p} \;,
\ee
in which the spinor and scalar components are, respectively~\cite{Volkov,Neville:1971uc},
\be\begin{split}
	K_p(\phi) &= 1+\frac{\fsl{k}\fsl{a}(\phi)}{2kp} \;, \\
\varphi_{p}(x) &= \exp -\bigg[ipx+\int\limits_0^\phi\frac{2pa-a^2}{2kp}\bigg] \;.
\end{split}
\ee
The commutators of the fermions modes are
\be
	\{b_\lambda(p),b^\dagger_{\lambda'}(p')\} = \{d_\lambda(p),d^\dagger_{\lambda'}(p')\} = \delta({\sf p},{\sf p}')\delta_{\lambda \lambda'} \;,
\ee
The lightfront QED Hamiltonian contains two groups of terms. The first is order one in the coupling $e$,
\be
	H_1=\frac{e}{2}\tint{x}\ \bar{\Psi} A_\mu \gamma^\mu\Psi \;,
\ee
(where $\ud{\sf x} = \ud^2x^\LCp\ud x^\LCm$) and the second group is order $e^2$. The latter contains the seagull and four-fermion terms, both particular to lightfront quantisation. We find that, ultimately, these terms do not contribute to the processes of interest here, and so we discard them from here on. (In terms of Feynman diagrams, one finds that the instantaneous parts of the fermion propagators do not contribute.) It is useful to note for later that, in a plane wave, the Lorentz momentum $\pi_\mu$ for a particle with initial momentum $p_\mu$ is 
\be
	\pi_\mu(\varphi;p) = p_\mu-a_\mu(\varphi)+\frac{2a(\varphi)p-a^2(\varphi)}{2kp}k_\mu \;.
\ee
\subsection{The one loop amplitude}
The one-loop amplitudes for helicity flip and non-flip between single photon states are readily written down in terms of the time-evolution operator and the $S$-matrix. We are also interested in expectation values of the electromagnetic fields of a coherent (probe) photon state, at non-asymptotic times. Because coherent states are eigenvectors of the photon annihilation operators the calculations for the expectation values and single photon amplitudes are very similar. The relations between them can be found in Section~\ref{SECT:NEW}. Working to one-loop, we find that all the objects of interest are built from the following lightfront time-ordered product, a transition amplitude between photon states, momentum $l_\mu$ and polarisation $\epsilon_\mu$ to momentum $l'_\mu$ and polarisation $\epsilon'_\mu$,
\be\label{general}
\mathcal{M}(x^\LCp) = -\bra{l',\epsilon'} \!\! \int\limits^{x^\LCp}_{-\infty} \!\ud y^\LCp \!\!\! \int\limits^{y^\LCp}_{-\infty} \!\ud z^\LCp H_1(y^\LCp) H_1(z^\LCp) \ket{l,\epsilon} \;.
\ee
We will briefly describe the computation of this object, and then relate it to the observables of interest. We need first to evaluate the action of the Hamiltonian on the incoming state,
\be
		H_1(z^\LCp) \ket{l,\epsilon} = H_1(z^\LCp)\epsilon(l)a^\dagger(l)\ket{0} \; .
\ee
The integral over position ${\sf x}$ in $H_1$ yields a delta function in three momentum components. Because all momenta have positive minus component, $q_\LCm>0$, and because there are no fermions in the incoming state, only terms like
\be
	H_1(z^\LCp)\epsilon(l)a^\dagger(l)\ket{0} \sim b^\dagger d^\dagger a a^\dagger(l) \ket{0} \; 
\ee
can contribute. The photon operators can then be commutated away. It follows that the only contributing term of the second factor of $H_1$ in (\ref{general}) can be $\sim b d a^\dagger$, and all the commutators can be performed immediately. One finds, after a standard calculation of the trace that $\mathcal{M} = \delta({\sf l}',{\sf l})iT_a$, where
\bw
\be\begin{split}\label{Ra}
iT_a = \frac{e^2}{kl} \int\limits^{kx}\!\ud\phi_2\! \int\limits^{\phi_2}\!\ud\phi_1 \!\int\!\ud{\sf p}\frac{\theta(l-p)_\LCm}{k(l-p)} e^{-i\int\limits_1^2\frac{l\pi}{k(l-p)}}
 \bigg( \bar{\epsilon}'\underset{(2}{\pi}\epsilon\underset{1)}{\pi} &+ \bigg[1-\frac{1}{2}\frac{kl^2}{kpk(l-p)}\bigg]\bar{\epsilon}'\underset{[2}{\pi}\epsilon\underset{1]}{\pi}+ \bar{\epsilon}'\epsilon\frac{kl^2}{4kp k(l-p)} (a_2-a_1)^2  \\
 &+\bar{\epsilon}'\epsilon\frac{kl}{2k(l-p)}(l\pi_2+l\pi_1)\bigg) \;,
\end{split}
\ee
%
in which $1,2$ indicate the arguments $\phi_1,\phi_2$ and bracketed subscripts mean (anti-)symmetrisation.  The integrals in (\ref{Ra}) are regulated using transverse dimensional regularisation~\cite{Casher:1976ae}. The field-dependent terms are UV finite, and renormalisation is performed by subtracting the divergent free-field contribution from the integrand of $T_a$, which we schematically write as $T_a-T_0$. After this subtraction, one can return to 3+1 dimensions. The renormalised amplitude $T_{\epsilon'\epsilon} = T_a - T_0$ is then UV finite and vanishes when the field vanishes, so that e.g.\ the full forward scattering amplitude becomes $\mathcal{F}=1$. From this result one can directly extract the refractive indices using (\ref{F1-n}), or take the asymptotic limit to obtain the scattering amplitudes.

However, $T_a-T_0$ in the form (\ref{Ra}) is cumbersome to evaluate numerically, and we would like to be able to perform as many of the integrals as possible analytically. We will present this for the asymptotic limit $T(\infty)$, which is a little simpler since e.g.\ the final term in (\ref{Ra}) vanishes immediately. That term is a total derivative,
\be
	(l\pi_2+l\pi_1) e^{...}=ik(l-p)(\partial_2-\partial_1)e^{...} \;,
\ee
and in the asymptotic limit gives a vanishing boundary contribution.

So, following the renormalisation above, we can perform the $p_\LCperp$ integrals with ease, since they are Gaussian~\cite{Dinu:2013hsd}. We also change variables from the ordered times to $\phi=(\phi_2+\phi_1)/2$ and $\theta=\phi_2-\phi_1$, and from $p_\LCm$ to the lightfront momentum fraction $s := p_\LCm/l_\LCm$. This leads to expressions containing the form (writing $c = 1/(2kl s(1-s))$)
\be\label{renorm1}
	\frac{e^{-ic\theta M^2}-e^{-ic\theta m^2}}{\theta^2} \;,
\ee
which come from renormalisation, i.e.\ from the subtraction of the free terms. In this form it is clear that the UV divergence would, without renormalisation, appear as a contact term in position space, at $\theta=0$~\cite{Dittrich:2000zu}. It is easily verified by expanding $M^2$ and (\ref{renorm1}) in powers of $\theta$ that subtracting the free part removes this $1/\theta^2$ divergence. It is helpful, especially for numerics, to rewrite such terms with a single exponent:
\be\label{renorm2}
	\int\limits_0^\infty\!\ud\theta\ \frac{e^{-ic\theta M^2}-e^{-ic\theta m^2}}{\theta^2} =-ic\int\limits_0^\infty\!\ud\theta\
	e^{-ic\theta M^2} \frac{M^2-m^2}{\theta M^2}\frac{\ud \theta M^2}{\ud\theta} \;,
\ee
as can be shown by integrating by parts (the boundary term vanishes), separating out the lowest order behaviour in $\theta$, and using the properties of the effective mass $M^2$~\cite{Kibble:1975vz, Hebenstreit:2010cc}. One is left at this stage with
\be\label{T-exact}
\begin{split}
T_{\epsilon'\epsilon}(\infty)=-\frac{\alpha}{2\pi}\frac{1}{kl}\int\limits_{-\infty}^\infty\ud\phi\int\limits_0^\infty\ud\theta\theta\int\limits_0^1\ud se^{-\frac{i}{2kl}\frac{\theta M^2}{s(1-s)}}\bigg(\epsilon'\epsilon & \bigg[\frac{M^2-m^2}{\theta^2M^2}\frac{\ud(\theta M^2)}{\ud\theta}+\frac{\langle a\rangle_\phi^2}{4s(1-s)}\bigg] \\
&-\frac{1}{2}\bar{A}_\phi A_\phi+2\bar{A}_\theta A_\theta+\bigg[1-\frac{1}{2s(1-s)}\bigg]\bar{A}_{[\phi}A_{\theta]}\bigg) \;,
\end{split}
\ee
\ew
where $A\equiv \epsilon\langle a\rangle$, $\bar{A}=\bar{\epsilon}'\langle a\rangle$ and subscript $\phi$ and $\theta$ indicates derivatives, e.g. $A_\theta \equiv \epsilon^\mu \partial_\theta \langle a_\mu \rangle$.  We now turn to the $s$-integral, noting that  the $s(1-s)$ factors in the exponents are typical of lightfront wavefunctions~\cite{Lepage, Brodsky:1997de,Heinzl:2000ht}.  This final momentum integral can also be performed analytically, using the results
\be\begin{split}\label{Bessel1}
	&\int\limits_0^1\!\frac{\ud s}{2s(1-s)}\ e^{\frac{-ix}{2s(1-s)}} = e^{-ix} K_0(ix) \;, \\
	&\int\limits_0^1\!\ud s\ e^{\frac{-ix}{2s(1-s)}} =i x e^{-ix} \big(K_1(ix)- K_0(ix) \big) \;,
\end{split}
\ee
which finally leads to the expression (\ref{T-resultat}) in the text. For the low energy limit we rescale the $\theta$ integral $\theta\to (kl/m^2)\theta$ and expand in $kl/m^2$. To lowest order the last two terms in (\ref{T-exact}) vanish. This expansion can be performed, if one wishes, before the $s$-integral is performed. Then the $\theta$ integral becomes
\be
	\int\limits_0^\infty\ud\theta\theta e^{-ib\theta}=-\frac{1}{b^2} \;,
\ee
after which the $s$ integral becomes trivial. We have presented the calculation for $T(\infty)$. However, for small $kl/m^2$ the amplitude has the same form for finite times, except that the $\phi$-integral extends to $kx$ rather than infinity. The result is as shown in (\ref{Low energy R}).

\section{Crossed field case}\label{APP:CROSSED}
Here we will show equivalence between (\ref{F1-resultat}) and known literature results in the case of a constant crossed field, for which the polarisation tensor is well known. Contracting that tensor with $\epsilon$'s should then yield the appropriate non-flip amplitude, in a crossed field.  Note that our results are both more general and more compact than the majority in the literature (and that we have already reproduced the known refractive indices). Demonstrating the equivalence between our results and others is a a tedious exercise in changing variables, in order to show the equivalence of two at-first-sight different integrals.

For this reason, we present only the simplest case, which is to assume $\epsilon a=0$. We start from (\ref{T-exact}), with the $s$-integral unevaluated, which is closer in form to the literature expressions. The non-flip amplitude is then (still for arbitrary plane waves)
\bw
\be\label{F1-resultat2}
\begin{split}
	T_{\epsilon'\epsilon}(\infty) = \frac{\alpha}{2\pi} \frac{1}{kl}
	\int\limits_{-\infty}^\infty\!\ud\phi \! 
	\int\limits_0^\infty\ud\theta\theta\!
	\int\limits_0^1\!\ud s\ 
	e^{-\frac{i}{2kl}\frac{\theta M^2}{s(1-s)}} \bigg[\frac{\langle a\rangle_\phi^2}{4s(1-s)} + \frac{\ud\theta M^2}{\ud\theta}\frac{M^2-m^2}{\theta^2M^2}\bigg] \;.
\end{split}
\ee
\ew
In a crossed field we have $a_\mu = m\mathcal{E}_\mu\phi$, taking $\omega=m$ for convenience, and $M^2 = m^2(1-\mathcal{E}^2\theta^2/12)$. (Recall that $-\mathcal{E}^2=E^2/E_S>0$.) In a crossed field, (\ref{F1-resultat2}) becomes independent of $\phi$, and the $\ud\phi$-integral contributes the infinite volume of lightfront time, which we can divide out. Now take the first term in (\ref{F1-resultat2}) and change variables as follows:
\be
	t := \frac{\theta}{2 b_0 s (1-s)} \;, \quad s:=\tfrac{1}{2}(1+v) \;.
\ee
Then the first term becomes
\be\label{usch1}
	\frac{-\alpha b_0}{16\pi}\bigg(\frac{E}{E_S}\bigg)^2\int\limits_{-1}^1\!\ud v\, (1-v^2) \! \int\limits_0^\infty\!\ud t\, t  \exp\big[-i t -i \tfrac{1}{48}L^2 t^3\big] \;,
\ee
where $L = b_0(1-v^2)E/E_S$. For the second term in (\ref{F1-resultat2}), we rewrite in a more convenient form, this time introducing an extra integral:
\be\label{F1-resultat3}
e^{-\frac{i}{2kl}\frac{\theta M^2}{s(1-s)}}-e^{-\frac{i}{2kl}\frac{\theta m^2}{s(1-s)}} \propto \int\limits_0^1\!\ud\eta\ e^{-\frac{i}{2kl}\frac{\theta M^2_\eta}{s(1-s)}} \;,
\ee
where $M^2_\eta = m^2 + \eta(M^2-m^2)$. Change variables $\theta\to t$ as above, then from $\eta \to \xi := s^2(1-s)^2\eta$. The integrand becomes independent of $s$, and that integral can be performed. After this, one can change variables again, $\xi \to v$ with $\xi=(1-v^2)^2/16$. (Note that these changes of variables must be defined `piecewise' in order to be invertible and properly cover the integration volume.) One obtains (\ref{usch1}) but with an additional factor of $-v^2/3$ in the integrand. Finally, adding the two terms together gives us
\bw
\be
	\frac{T_{\epsilon\epsilon}(\infty)}{\int\!\ud\phi} = \frac{-\alpha b_0}{16\pi}\bigg(\frac{E}{E_S}\bigg)^2\int\limits_{-1}^1\!\ud v\, (1-v^2)(1-\tfrac{1}{3}v^2) \int\limits_0^\infty\!\ud t\, t\,  \exp\big[-i t -i \tfrac{1}{48}L^2 t^3\big] \;,
\ee
\ew
in which we have the same integral as in equation 4.26 in \cite{Shore:2007um} and in equation (4) in~\cite{Narozhny:1968}) (contracted with $\epsilon$ where $\epsilon a=0$). 

\section{Numerical methods}\label{APP-C}
The final state and loop momenta in our amplitudes have been performed analytically. What remains in e.g.\~(\ref{T-exact}) are the $\{\theta,\phi\}$ integrals over lightfront times, which come from the two vertices in Fig.~\ref{FIG:LOOP}. For constant or monochromatic fields one can make some further analytic progress with these integrals, but for general fields one must turn to numerical methods.

For small $b_0$ and/or large $a_0$ a direct $\theta$ integration is difficult due to the integrand oscillating rapidly. Specialised numerical quadratures for integrals involving trig factors have been
employed, by separating the oscillatory factor $e^{-2i\frac{\theta M^{2}}{kl}}$ from the special functions $\mathcal{I}_{j}$ (the remaining factor being well behaved) and numerically
changing variable from $\theta \to \Theta :=\theta M^2/m^2$ in
order to make oscillations uniform. (For very small $\theta$ and large $\Theta/b_0$, perturbative and asymptotic expansions can help to maintain precision.) It is easily verified that
\be\label{D1}
	\frac{\partial\Theta}{\partial\theta} = 1 - \frac{(  a_1+a_2-2\langle a\rangle )
^{2} + (a_2 - a_1)^2}{4m^2}\geq1 \;,
\ee
so the change of variables is well-defined. (Subscripts once again denote the arguments $\phi_2 = \phi+\theta/2$, $\phi_1=\phi-\theta/2$.) We note also that
\be\label{D2}
	\frac{\partial\Theta}{\partial\phi} = -\frac{(  a_1 + a_2 - 2\langle a\rangle)(
a_2-a_1)  }{m^{2}}
\ee
For $a_0^2 \gg b_0$ the main contribution to our integrals comes from the vicinity of points where
\be\label{condition}
	a_1= a_2=\langle a\rangle \;,
\ee
so that (\ref{D1}) is minimised and (\ref{D2}) vanishes. The dominant contribution is then obtained by expanding in $\theta$ about $\theta=0$ (for any $\phi$), this giving the `locally constant' result based on crossed fields.

Other points $\{\theta_{i},\phi_{i}\}$ such that the derivatives are minimised generally exist only
for linearly polarised backgrounds. These points give a correction which, for moderate $a_0$, oscillates as a function of $b_{0}$, as can be seen in for example Fig.~\ref{FIG:SCALING}. It is worth noting that these oscillations are linked to the presence of the effective mass, as follows. To illustrate, let the field be monochromatic, $a_\LCperp=\{a_0 m \sin\phi,0\}$. The condition (\ref{condition}) is obeyed at (amongst others) the points $\{\phi_i,\theta_i\} = \{\{0,\pi\},2r\pi\}$, $r\in\mathbb{Z}^+$ (recall that $\phi$--integrals are restricted to one cycle for monochromatic plots).  At these points, second derivatives of $\Theta$ also vanish and, for example,
\be
	\Theta(  \phi_i,\theta_i)  =\theta_i(  1+ \tfrac{1}{2} a_0^2) \;,
\ee
i.e.\ Kibble's mass $M^2$ becomes equal to its asymptotic value
\be
	M^2 \to m_*^2 = m^2 (1+ \tfrac{1}{2} a_0^2)  \;,
\ee
which we recognise as the well-known effective mass in a laser pulse~\cite{Kibble:1975vz,Kibble:1965zz,Harvey:2012ie}.

\end{document}